\documentclass[a4paper,11pt]{article}
\usepackage{epsfig}
\usepackage{amsmath}
\usepackage{axodraw}
\usepackage{graphicx}

\numberwithin{equation}{section}

\begin{document}
\title{
Mixing-induced CP asymmetry
in ${B\to K^*\gamma}$ decays with
perturbative QCD approach}

\author{
M.~Matsumori$^{a}$\footnote{Electronic address:
mika@eken.phys.nagoya-u.ac.jp} 
~and
A.~I.~Sanda$^{a,b}$
\footnote{Electronic address: sanda@eken.phys.nagoya-u.ac.jp}
\\
{ \small $^a$ Department of Physics, Nagoya University, Chikusa-ku, Furo-cho,
    Nagoya, 464-8602, Japan}\\
{\small $^b$ 
Kanazawa University, 3-27-1 Rokukakubashi, Kanazawa-ku, Yokohama-shi,
Yokohama, Japan}
}

\date{}
\maketitle
\begin{abstract}
The  mixing-induced CP asymmetries
in ${B\to K^*\gamma \to K_S\pi^0\gamma}$ and 
${B\to K^*\gamma\to K_L\pi^0\gamma}$
are expected to be small
within the standard model.
So they are among the most promising decay modes
to test the standard model.
In this paper, we compute the mixing-induced CP asymmetries
in these decay modes, within the framework of 
the standard model, using   perturbative QCD,
and our conclusion is 
${S_{K_S \pi^0\gamma}=-S_{ K_L \pi^0\gamma}=-(3.5\pm 1.7)\times 10^{-2}}$.
\end{abstract}
\section{Introduction}
The standard model (SM), which includes the
Kobayashi-Maskawa (KM) scheme for CP violation \cite{Kobayashi:1973fv},
is being tested
by determining the three sides and three angles of the unitary
triangle.
The observations of the  CP asymmetries
in 
${B\to J/\psi K_S}$ \cite{Abe:2003yu,Raven:2003gs},
 ${B\to \pi\pi}$ \cite{Abe:2005dz,Bevan:2004ht},
and ${B\to DK}$ \cite{Abe:2004gu,Aubert:2005iz}
decays 
show that a major contribution to CP violation
comes from the KM scheme.
We know that the KM scheme is not the whole story.
Baryogenesis cannot be explained by this scheme.
To understand this, we
must search for new physics.
It is therefore important to consider 
all possible CP violating phenomena.

Among many
${B}$ meson decay modes,
the radiative decay ${B\to K^*\gamma}$ 
attracted much attention because
it is a
flavor-changing-neutral-current  
process, which occurs only through  quantum corrections.
So it is expected
to be 
sensitive to the
physics beyond the SM.
This decay has been seen, and 
first 
experimental results on CP asymmetry and isospin breaking 
effects have been investigated
\cite{Nakao:2004th, Aubert:2004te}.
This has also been investigated within the framework of 
pertuyrbative QCD (pQCD)
\cite{Li:1998tk,Keum:2004is}.
We expect the
experimental situation to improve
in the near future.

\begin{figure}
 \begin{center}
\includegraphics[width=5cm]{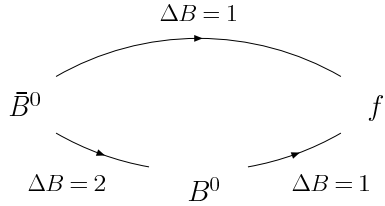}
\caption{
The mixing-induced CP asymmetry is caused by the
interference between ${\bar{B}^0\to f}$
and ${\bar{B}^0\to B^0\to f}$ decay amplitudes.}
\label{M}
\end{center}
\end{figure}

\begin{figure}
 \begin{center}
\includegraphics[width=6cm]{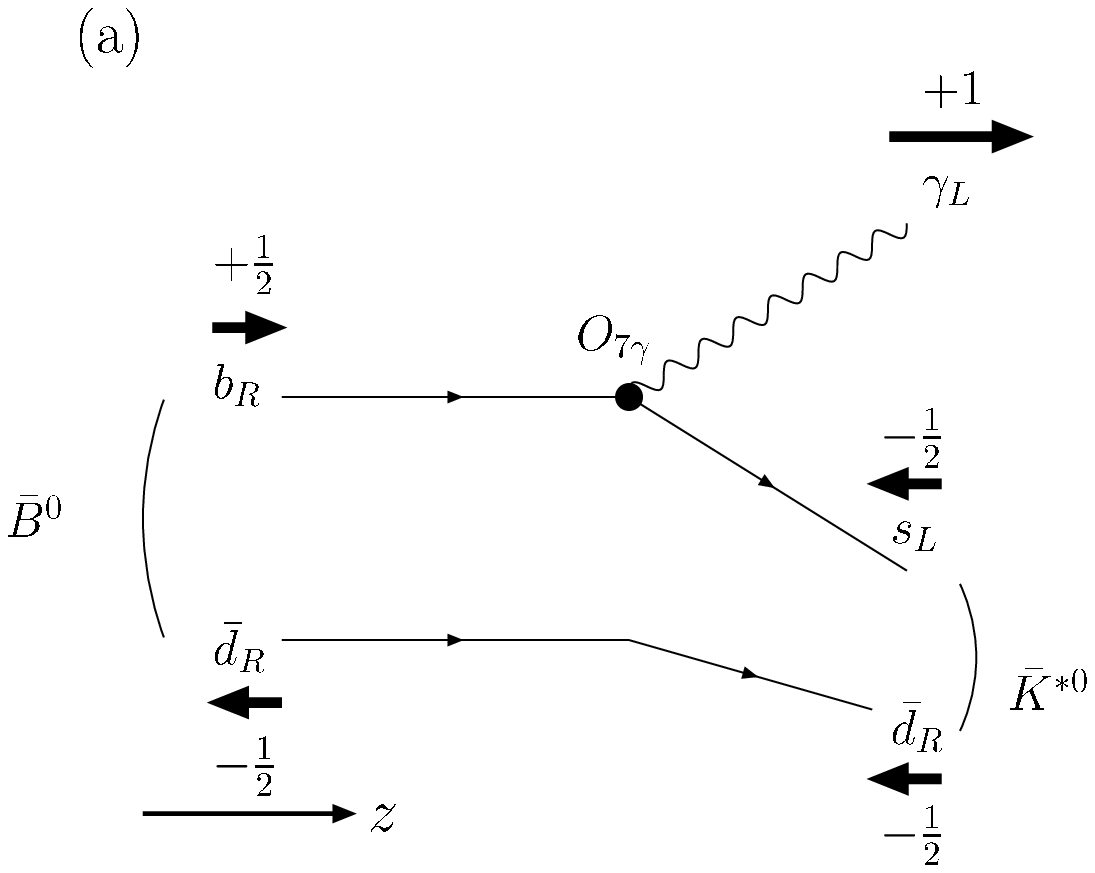}
\hspace{1cm}
\includegraphics[width=6cm]{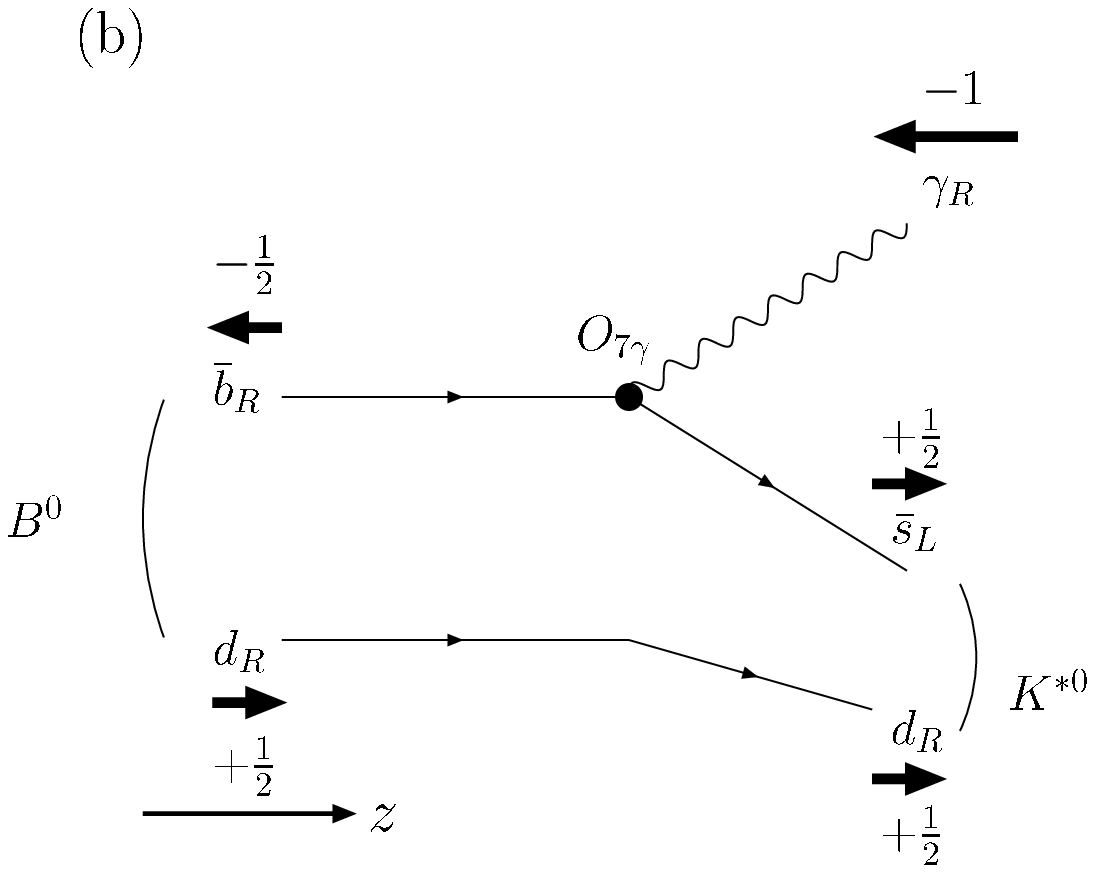}
\caption{
Figures (a), and (b) show the major decay modes and
the helicity configurations of the radiative
decays of ${b_R}$, and ${\bar{b}_R}$, respectively.
}
\label{helicity}
\end{center}
\end{figure}

There is a good reason why
the mixing-induced CP asymmetry in ${B\to K^*\gamma}$
is expected to be small within the SM.
If ${B^0}$ and ${\bar{B}^0}$ mesons can decay
into a common final state ${f}$, the mixing-induced
CP asymmetry is generated by the interference 
between ${\bar{B}^0\to f}$ and 
${\bar{B}^0\to B^0\to f}$
decay amplitudes as shown  in Fig.\ref{M}.
For
${B\to K^*\gamma}$ decay, the candidates
for ${f}$ are ${K_S\pi^0\gamma}$ and ${K_L\pi^0\gamma}$.
Throughout this paper, whenever we write
${B^0\to K_{(S,L)}\pi^0\gamma}$,
we mean ${K_{(S,L)}\pi^0 \gamma}$ to come from
${B^0\to K^{*0}\gamma\to K_{(S,L)}\pi^0\gamma}$.
The continuum background ${K_{(S,L)}\pi^0}$
which is not from the ${K^{*0}}$ decay
can be subtracted.
While a detailed experimental study is necessary
to understood  the error on the asymmetry
coming from this background, we guess
that the error is much smaller than the
theoretical error included in our result.
The ${\bar{B}^0}$ meson
dominantly decays
into a photon with left-handed chirality
as shown in Fig.\ref{helicity}(a), because
the chirality is automatically determined by the vertex structure.
In order to generate the asymmetry,
the amplitudes representing the two paths shown in Fig.\ref{M}
must interfere. Both ${B^0}$ and ${\bar{B}^0}$ must
decay to the same final state ${f}$.
However,  the ${B^0}$ meson
dominantly decays into a photon with right-handed chirality
as shown in Fig.\ref{helicity}(b).
That is,
the dominant contributions of 
${\bar{B}^0}$ and ${B^0}$ meson decays have
different photon chiralities, and
they
cannot interfere. 
The amplitude which can interfere,
${A(B^0\to K_{(S,L)}\pi^0\gamma_L)}$,
is suppressed by the factor of ${m_s/m_b}$
compared to
${A(B^0\to K_{(S,L)}\pi^0\gamma_R)}$;
thus the asymmetry which is generated by
${B^0 \leftrightarrow \bar{B}^0}$ mixing 
is suppressed by ${m_s/m_b}$ 
\cite{Atwood:1997zr}.
If the experimental data show a large mixing-induced
CP asymmetry, 
much larger than ${10\%}$, for example,
it directly indicates that there exist
a new physics beyond the SM.
So we expect that the mixing-induced CP asymmetry in
the ${B\to K^*\gamma}$ decay mode 
is a crucial decay mode to test the SM.
The experimental data for mixing-induced
CP asymmetry in ${B\to K^*\gamma}$ are given
by Belle and BaBar as follows:
\begin{equation*}
S_{K^*\gamma\to K_S\pi^0\gamma}^{\mbox{\scriptsize{ex}}}=
\begin{cases}
-0.79^{+0.63}_{-0.50}\pm 0.10 & \mbox{\cite{Ushiroda:2005sb,Abe:2004xp}},\\
-0.21\pm 0.40 \pm 0.05 & \mbox{\cite{Aubert:2005bu}}.
\end{cases}
\end{equation*}

This paper is organized as follows:
in Sec.\ref{Bdecay}, we want to review
the ${B}$ meson decay and mixing-induced CP
asymmetry, and
in Sec.\ref{formula}, we show 
decay amplitudes for ${B\to K^*\gamma}$ decay
which is classified into
left-handed or right-handed photon chiralities.
We show the numerical results in Sec.\ref{Numerical}, 
and in  Sec.\ref{conclusion},
we present our conclusion on the
mixing-induced CP asymmetry
in ${B\to K^*\gamma}$ decay. 

\section{${B}$ meson decay}
\label{Bdecay}
\subsection{Mixing-induced CP asymmetry}
Here we want to review the mixing-induced CP asymmetry
in the  neutral ${B}$ meson decay system 
\cite{Carter:1980hr,Carter:1980tk,Bigi:1981qs}.
There are two neutral ${B}$ mesons, ${B^0}$ and ${\bar{B}^0}$,
and if we turn off the weak interaction,
${B^0}$ and ${\bar{B}^0}$ mesons are independent of each other,
and the transition between ${B^0}$ and ${\bar{B}^0}$ mesons
does not occur.
But if the weak interaction is once turn on, ${B^0}$ and ${\bar{B}^0}$
mesons change each other through the common mediated 
states ${f}$.
The time-dependent CP asymmetry
is given as follows:
\begin{eqnarray}
A_{\mbox{\scriptsize{cp}}}(t)&=&\frac{\Gamma(\bar{B}^0(t)\to f )-
\Gamma(B^0(t)\to f)}
{\Gamma(\bar{B}^0(t)\to f )+\Gamma(B^0(t)\to f)}
=A\cos{\Delta Mt}
+S
\sin{\Delta Mt},
\end{eqnarray}
\begin{eqnarray}
A=-\frac{1-|\bar{\rho}|^2}{1+|\bar{\rho}|^2},~~~~~~~~
S=\frac{2\mbox{Im}\left[\frac{q}{p}
%e^{-2i\phi_1}
\bar{\rho}\right]}{1+|\bar{\rho}|^2}~,
\end{eqnarray}
where
\begin{eqnarray}
A(f)=\langle f | H_{\mbox{\scriptsize{eff}}}|B^0\rangle,\hspace{8mm}
\bar{A}(f)=\langle f | H_{\mbox{\scriptsize{eff}}}
|\bar{B}^0\rangle,\hspace{8mm}
\bar{\rho}=
\frac{\bar{A}(f)}{A(f)}~.
\end{eqnarray}
Here ${S}$ is called  the mixing-induced CP asymmetry,
and in the ${B\to J/\psi K_S}$ case, for example,
${S=\sin{2\phi_1}}$.
We can test the SM by determining the three angles
from
${\phi_1}$ to ${\phi_3}$ 
by investigating the mixing-induced CP
asymmetries in some decay channels.

\subsection{Mixing-induced CP asymmetry in ${B^0\to K_S\pi^0\gamma}$
and ${B^0\to K_L\pi^0\gamma}$}
Next we want to discuss  the mixing-induced CP asymmetry
in ${B\to K^*\gamma}$ decay.
In order to examine the 
CP asymmetry,
we have to look for the common final 
states between ${B^0}$
and ${\bar{B}^0}$ meson decays.
The candidates in ${B\to K^*\gamma}$ decay 
are ${B\to K_S\pi^0\gamma}$ and ${B\to K_L\pi^0\gamma}$; ${K^*}$ meson
decays into  ${K^0}$ and ${\pi^0}$, and ${K^0}$ meson
goes to ${K_S}$ or ${K_L}$.

The effective Hamiltonian which
induces the flavor-changing ${b\rightarrow s \gamma}$
transition is given by \cite{Buchalla:1995vs}
\begin{eqnarray}
H_{\mbox{\scriptsize{eff}}}(\Delta B=1 )&=&\frac{G_F}{\sqrt 2}
\Big[
\sum_{q=u,c} V_{qb}V_{qs}^*\left\{
C_1(\mu)O_1^{(q)}(\mu)+C_2(\mu)O_2^{(q)}(\mu)\right\}\nonumber\\
&&\hspace{5mm}-V_{tb}V_{ts}^*\left\{
\sum_{i=3\sim 6}C_{i}O_{i}(\mu)+C_{7\gamma}O_{7\gamma}(\mu)
+C_{8g}(\mu)O_{8g}(\mu)\right\}
\Big]+\mbox{h.c.},\nonumber\\
\end{eqnarray}
where $C_i$'s are Wilson coefficients, and $O_i$'s are local operators
which are given by
\begin{eqnarray}
&&O_1^{(q)}=(\bar{s}_i q_j)_{V-A}(\bar{q}_j b_i)_{V-A}
,\hspace{2.6cm}O_2^{(q)}=(\bar{s}_i q_i)_{V-A}(\bar{q}_j b_j)_{V-A},\nonumber\\
&&O_3=(\bar{s}_i b_i)_{V-A}\sum_q
(\bar{q}_j q_j)_{V-A}
,\hspace{2.2cm}O_4=(\bar{s}_i b_j)_{V-A}\sum_q
(\bar{q}_j q_i)_{V-A},\nonumber\\
&&O_5=(\bar{s}_i b_i)_{V-A}\sum_q
(\bar{q}_j q_j)_{V+A}
,\hspace{2.2cm}O_6=(\bar{s}_i b_j)_{V-A}\sum_q
(\bar{q}_j q_i)_{V+A},
\\
&&O_{7\gamma}=\frac{e}{4{\pi}^2}\bar{s}_i\sigma^{\mu \nu}(m_sP_L
 +m_bP_R)b_iF_{\mu \nu},\hspace{7mm}
O_{8g}=\frac{g}{4{\pi}^2}\bar{s}_i\sigma^{\mu \nu}(m_sP_L
 +m_bP_R)T_{ij}^ab_jG_{\mu \nu}^a.\nonumber
\label{operator}
\end{eqnarray}
Here we set ${P_L^R=(1\pm\gamma^5)/2}$ and
${(\bar{q}_i q_i)_{V\mp A}}$ means 
${2\bar{q}_i \gamma^{\mu}P^L_R~ q_i}$,
where ${i}$ and ${j}$ are color indexes.

\begin{figure}
 \begin{center}
\includegraphics[width=8cm]{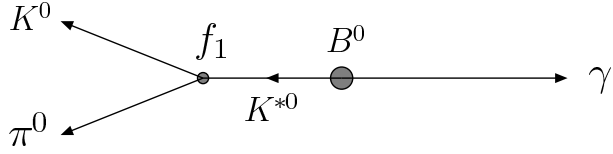}
\caption{The figure shows the ${B^0\to K^{*0}\gamma\to K^0\pi^0\gamma}$ 
decay mode.
The ${K^{*0}\to K^0\pi^0}$ process possess the complex factor
${f_1}$ which is common in both  ${K^{*0}\to K^0\pi^0}$ and
 ${\bar{K}^{*0}\to \bar{K}^0\pi^0}$ because the strong interaction
conserves the C and P symmetries.}
\label{common}
\end{center}
\end{figure}

We consider
the decay amplitudes ${B\to K_{S}\pi^0\gamma}$ 
and ${B\to K_{L}\pi^0\gamma}$. 
The decay amplitude ${A(B\to K_{S,L}\pi^0\gamma)}$ can be
extracted as follows:
first the ${B}$ meson decays into ${K^*}$ and ${\gamma}$
by the weak interaction.
%When we consider 
The decay amplitudes %in Eqs.(\ref{27}) 
caused by
${O_{7\gamma}}$ operator can be expressed as
\begin{eqnarray}
A_L&\equiv &A(B^0\to K^{*0}\gamma_L)=
Fm_s
\langle K^{*0}\gamma_L |\bar{b}\sigma^{\mu\nu}(1+\gamma^5)s F_{\mu\nu}|
B^0\rangle,\nonumber\\
A_R&\equiv & A(B^0\to K^{*0}\gamma_R)=
Fm_b
\langle K^{*0}\gamma_R |\bar{b}\sigma^{\mu\nu}(1-\gamma^5)s F_{\mu\nu}|
B^0\rangle,\nonumber\\
\bar{A}_L&\equiv &A(\bar{B}^0\to \bar{K}^{*0}\gamma_L)=
F^*m_b
\langle \bar{K}^{*0}\gamma_L |\bar{s}\sigma^{\mu\nu}(1+\gamma^5)b F_{\mu\nu}|
\bar{B}^0\rangle,\label{2.25}\\
\bar{A}_R&\equiv &A(\bar{B}^0\to \bar{K}^{*0}\gamma_R)=
F^*m_s
\langle \bar{K}^{*0}\gamma_R |\bar{s}\sigma^{\mu\nu}(1-\gamma^5)b F_{\mu\nu}|
\bar{B}^0\rangle,\nonumber
\end{eqnarray}
where we set the common factor;
\begin{eqnarray}
F\equiv V_{tb}^*V_{ts}\frac{e}{8\pi^2}.
\end{eqnarray}
At the  next stage, ${K^*}$ meson decays into ${K^0}$ and ${\pi^0}$
mesons through the strong interaction.
The strong interaction conserves the
C and P symmetries,
then we can express the decay process  
by a common  complex factor ${f_1\equiv A(K^{*0}\to K^0\pi^0)=
A(\bar{K}^{*0}\to \bar{K}^0\pi^0)}$
as shown in Fig.\ref{common}. 
Finally, we consider the process in which the ${K^0}$ meson
goes to ${K_S}$ or ${K_L}$.
Here the indexes ${S}$, and ${L}$, express the ``short'', and ``long''
lifetimes, of two ${K}$ mesons, respectively,
and denote the amplitudes for ${K^0\to K_{S,L}}$ and 
${\bar{K}^0\to K_{S,L}}$ as follows:
\begin{eqnarray}
&&A(K^0\to K_S)=f_S,\hspace{1cm}A(K^0\to K_L)=f_L,\nonumber\\
&&A(\bar{K}^0\to K_S)=f_S,\hspace{1cm}A(\bar{K}^0\to K_L)=-f_L.
\end{eqnarray}
Then by
denoting ${F_S=f_1f_S}$ and ${F_L=f_1f_L}$,
 the decay amplitudes for ${B\to K_S\pi^0\gamma}$ and 
${B\to K_L\pi^0\gamma}$
can be expressed as follows:
\begin{eqnarray}
&&A(B^0\to K_S\pi^0\gamma_L)=
F_S A_L,\hspace{1cm}%\nonumber\\
A(B^0\to K_S\pi^0\gamma_R)=
F_S A_R,\label{25}\\
&&A(\bar{B}^0\to K_S\pi^0\gamma_L)=
F_S\bar{A}_L,\hspace{1cm}%\label{25}\\
A(\bar{B}^0\to K_S\pi^0\gamma_R)=
F_S\bar{A}_R,\nonumber\\
\nonumber\\
&&A(B^0\to K_L\pi^0\gamma_L)=
F_L A_L,\hspace{1cm}%\nonumber\\
A(B^0\to K_L\pi^0\gamma_R)=
F_L A_R,\label{26}\\
&&A(\bar{B}^0\to K_L\pi^0\gamma_L)=
-F_L\bar{A}_L,\hspace{0.7cm}
A(\bar{B}^0\to K_L\pi^0\gamma_R)=
-F_L\bar{A}_R.\nonumber
\end{eqnarray}
Thus the time-dependent CP asymmetries become:
\begin{eqnarray}
A_{\mbox{\scriptsize{cp}}}
(B^0\to K_S\pi^0\gamma)&=&\frac{\Gamma(\bar{B}^0(t)\to K_S\pi^0\gamma )
-\Gamma(B^0(t)\to K_S\pi^0\gamma)}
{\Gamma(\bar{B}^0(t)\to K_S\pi^0\gamma )+\Gamma(B^0(t)\to
K_S\pi^0\gamma)}
\nonumber\\
&=&\frac{Y}{X}\cos{\Delta Mt}+\frac{Z}{X}\sin{\Delta Mt}\label{35},
\end{eqnarray}
\begin{eqnarray}
A_{\mbox{\scriptsize{cp}}}
(B^0\to K_L\pi^0\gamma)&=&\frac{\Gamma(\bar{B}^0(t)\to K_L\pi^0\gamma )
-\Gamma(B^0(t)\to K_L\pi^0\gamma)}
{\Gamma(\bar{B}^0(t)\to K_L\pi^0\gamma )+\Gamma(B^0(t)\to
K_L\pi^0\gamma)}
\nonumber\\
&=&\frac{Y}{X}\cos{\Delta Mt}-\frac{Z}{X}\sin{\Delta Mt},\label{36}
\end{eqnarray}
where we define
\begin{eqnarray}
X&=&|\bar{A}_L|^2+|\bar{A}_R|^2+|A_L|^2+|A_R|^2,\nonumber\\
Y&=&|\bar{A}_L|^2+|\bar{A}_R|^2-(|A_L|^2+|A_R|^2),\\
Z&=&2\mbox{Im}\left[e^{-2i\phi_1}(\bar{A}_LA_L^*+\bar{A}_RA_R^*)\right],\nonumber
\end{eqnarray}
and we can see that the signs of
the mixing-induced CP asymmetry 
between ${B\to K_S\pi^0\gamma}$
and ${B\to K_L\pi^0\gamma}$,
which are proportional to
${\sin{\Delta Mt}}$,
are opposite.
The mixing-induced CP asymmetries
are caused by the  interferences
between ${\bar{A}_L}$ and ${A_L}$,
${\bar{A}_R}$ and ${A_R}$.

The form factors in Eqs.(\ref{2.25}) are related by
C, P,
and CP symmetries.
We set the one form factor as the standard like 
${\langle K^{*0}\gamma_L |\bar{b}\sigma^{\mu\nu}(1+\gamma^5)s F_{\mu\nu}|
B^0\rangle \equiv F_1}$, the other
form factors can be expressed as % follows:
\begin{eqnarray}
\langle K^{*0}\gamma_R |\bar{b}\sigma^{\mu\nu}(1-\gamma^5)s F_{\mu\nu}|
B^0\rangle
&=&-
\langle K^{*0}\gamma_L |\mbox{P}^{\dagger}\mbox{P}
\left(\bar{b}\sigma^{\mu\nu}(1+\gamma^5)s F_{\mu\nu}\right)
\mbox{P}^{\dagger}\mbox{P}|
B^0\rangle =-F_1,\nonumber\\
\langle \bar{K}^{*0}\gamma_L |\bar{s}\sigma^{\mu\nu}(1+\gamma^5)b F_{\mu\nu}|
\bar{B}^0\rangle &=& 
\langle K^{*0}\gamma_L |\mbox{C}^{\dagger}\mbox{C}
\left(\bar{b}\sigma^{\mu\nu}(1+\gamma^5)s F_{\mu\nu}\right)
\mbox{C}^{\dagger}\mbox{C}|
B^0\rangle=F_1\\
\langle \bar{K}^{*0}\gamma_R |\bar{s}\sigma^{\mu\nu}(1-\gamma^5)b F_{\mu\nu}|
\bar{B}^0\rangle &=& -
\langle K^{*0}\gamma_L |\mbox{CP}^{\dagger}\mbox{CP}
\left(
\bar{b}\sigma^{\mu\nu}(1+\gamma^5)s F_{\mu\nu}\right)
\mbox{CP}^{\dagger}\mbox{CP}|
B^0\rangle=-F_1.\nonumber
\end{eqnarray}
Thus, Eqs.(\ref{2.25}) can be expressed as
\begin{eqnarray}
\label{2.28}
A_L=m_sF F_1,\hspace{8mm}
A_R=-m_bFF_1,\hspace{8mm}
\bar{A}_L=m_bF^*F_1,\hspace{8mm}
\bar{A}_R=-m_sF^*F_1,
\end{eqnarray}
and the mixing-induced CP asymmetries in 
Eqs.(\ref{35}) and (\ref{36}) become %as follows:
\begin{eqnarray}
S_{K_S\pi^0\gamma}=-S_{K_L\pi^0\gamma}=-2\frac{m_s}{m_b}\sin{2\phi_1}.
\end{eqnarray}
That is, the mixing-induced CP asymmetries are predicted
to be small within the SM roughly
by the factor of ${m_s/m_b}$
 as pointed out by \cite{Atwood:1997zr},
because
${A_L}$ and ${\bar{A}_R}$ are suppressed
compared to ${A_R}$ and ${\bar{A}_L}$.

However in the near future,
the mixing-induced CP asymmetry of ${B\to K^*\gamma}$
will become one of the most important  tests
for the SM,
so it is very important to estimate the CP asymmetry
more accurately by taking into account the 
small contributions
compared to the dominant contribution caused in the ${O_{7\gamma}}$
operator.
Cognizant of this point,
we also consider 
chromomagnetic penguin ${(O_{8g})}$,
the QCD annihilation ${(O_3\sim O_6)}$,
${u}$ and ${c}$ loop contributions,
and also the long distance effects.
Then we compute 
the mixing-induced CP asymmetry
in ${B\to K_S\pi^0\gamma}$ and
${B\to K_L\pi^0\gamma}$ decays within the SM,
and compare the experimental data given by
\cite{Ushiroda:2005sb,Abe:2004xp,Aubert:2005bu}.

\section{Factorization formula}
\label{formula}

The decay amplitudes can be factorized into
scalar ${(M^S)}$ and pseudoscalar ${(M^P)}$
components as
\begin{eqnarray}
M=(\epsilon_{\gamma}\cdot \epsilon_{K^*})M^S+i\epsilon_{\mu\nu+-}
\epsilon_{\gamma}^{\mu}\epsilon_{K^*}^{\nu}M^P
\label{3.1}
\end{eqnarray}
where ${\epsilon_{\gamma}}$, and ${\epsilon_{K^*}}$ are transverse
polarization vectors for ${\gamma}$, and ${K^*}$ mesons, respectively.
In the ${B\to K^*\gamma}$ decay mode,
the emitted photon is a real photon,
so the chirality of the photon is constrained to be left-handed
or right-handed,
because
the longitudinally polarized photon is forbidden by the gauge invariance.
Then the polarization vector of the photon
can be expressed as
\begin{eqnarray}
{\epsilon_{\gamma}}^L=\left(0,0,\frac{1}{\sqrt{2}}(1,i)\right),~~~~~~~
{\epsilon_{\gamma}}^R=\left(0,0,\frac{1}{\sqrt{2}}(1,-i)\right),
\label{45}
\end{eqnarray}
and polarization of the ${K^*}$ meson is uniquely determined 
by the helicity conservation of the
spinless ${B}$ meson decay as
\begin{eqnarray}
{\epsilon_{K^*}}^R=\left(0,0,\frac{1}{\sqrt{2}}(1,-i)\right),~~~~~~~
{\epsilon_{K^*}}^L=\left(0,0,\frac{1}{\sqrt{2}}(1,i)\right).
\label{46}
\end{eqnarray}
The detailed expressions for the ${B\to K^*\gamma}$ decay amplitudes
with the pQCD approach are in \cite{Keum:2004is},
then here we want to show the
way how to divide the ${{B}}$ meson  amplitudes into
the ones with left-handed or right-handed photon chiralities.

\subsection{${O_{7\gamma}}$ contribution}
\begin{figure}
\begin{center}
\includegraphics[width=4.2cm]{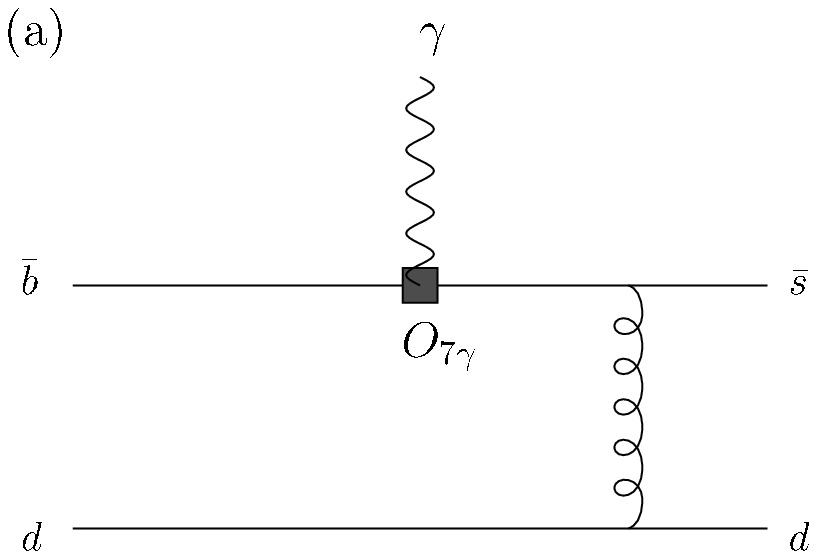}
\hspace{1cm}
\includegraphics[width=4.2cm]{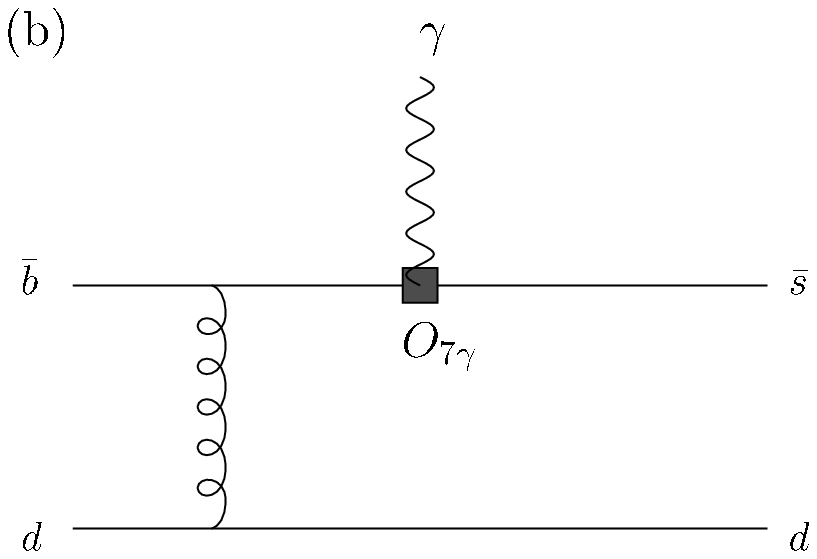}
  \end{center}
 \caption{The figure shows the ${B^0\to K^{*0}\gamma}$
 decay contribution caused by ${O_{7\gamma}}$ operator.
 The photon is emitted from the operator, 
 and the hard gluon exchange enable to form the
 fast moving ${K^*}$ meson.}
\label{O7}
\end{figure}

The decay amplitudes
caused by ${O_{7\gamma}}$  components which are proportional
to ${m_b}$ as shown in Fig.\ref{O7}
can be expressed as 
Eqs.(14) and (15) in  \cite{Keum:2004is}:
\begin{eqnarray}
&&M_{7\gamma}^{S (a)}(m_b)=-M_{7\gamma}^{P (a)}(m_b)
=r_bM_{7\gamma}^{S(a)},\\
\label{3.14}
&&M_{7\gamma}^{S(b)}(m_b)=-M_{7\gamma}^{P(b)}(m_b)
=r_bM_{7\gamma}^{S(b)},
\end{eqnarray}
where we define 
${r_q=m_q/M_B}$.
In this time, we distinctively express the quark mass ${m_b}$
which comes from the ${O_{7\gamma}}$ operator from the meson mass
${M_B}$.

Here we want to concentrate on the characteristic
relationship between the scalar and pseudo-scalar components as
${M_{7\gamma}^S(m_b)=-M_{7\gamma}^P(m_b)}$.
If we write down the decay amplitude as in the form of
Eq.(\ref{3.1}), we can factor out the common factor
as
\begin{eqnarray}
M=M_{7\gamma}^{S}(m_b)\left[
(\epsilon_{\gamma}\cdot \epsilon_{K^*})-i\epsilon_{\mu\nu +-}
\epsilon_{\gamma}^{\mu}\epsilon_{K^*}^{\nu}
\right].
\label{2}
\end{eqnarray}
When we enter  Eqs.(\ref{45}) and (\ref{46}) into
Eq.(\ref{2}), the combination ${{\epsilon_{\gamma}}^R}$
and ${{\epsilon_{K^*}}^L}$ only survives.
That is, the chiralities of the photon and ${K^*}$ meson
are automatically determined,
and in the ${B}$ meson decay caused by the component of 
${O_{7\gamma}}$ operator which is proportional to ${m_b}$, the
emitted photon has necessary right-handed chirality.

The decay amplitude caused by the
${O_{7\gamma}}$ operator component
which is proportional to ${m_s}$, 
on the other hand, 
becomes as follows:
\begin{eqnarray}
&&M_{7\gamma}^{S (a)}(m_s)=M_{7\gamma}^{P (a)}(m_s)=-r_sM_{7\gamma}^{S (a)},\\
&&M_{7\gamma}^{S(b)}(m_s)=M_{7\gamma}^{P(b)}(m_s)
=-r_sM_{7\gamma}^{S(b)}.
\end{eqnarray}
The characteristic relation ${M_{7\gamma}^S(m_s)=M_{7\gamma}^P(m_s)}$
implies that the combination ${\epsilon_{\gamma}^L}$ and 
${\epsilon_{K^*}^R}$ only survives.
It indicates that in the ${B}$ meson decay caused by the
component of the ${O_{7\gamma}}$ operator which is proportional
to ${m_s}$, the photon has left-handed chirality.

Furthermore, we investigate
the ${\bar{B}}$ meson decay.
The amplitude for ${\bar{B}}$ meson decay
can be extracted from  the hermite conjugate operator
for the ${B}$ meson decay:
\begin{eqnarray}
&&\bar{M}_{7\gamma}^{S (a)}(m_b)=\bar{M}_{7\gamma}^{P (a)}(m_b)
=-r_b\frac{\xi_t^*}{\xi_t}M_{7\gamma}^{S (a)},\\
&&\bar{M}_{7\gamma}^{S(b)}(m_b)=\bar{M}_{7\gamma}^{P(b)}(m_b)
=-r_b\frac{\xi_t^*}{\xi_t}M_{7\gamma}^{S (b)},
\end{eqnarray}
\begin{eqnarray}
&&\bar{M}_{7\gamma}^{S (a)}(m_s)=-\bar{M}_{7\gamma}^{P (a)}(m_s)
=r_s\frac{\xi_t^*}{\xi_t}M_{7\gamma}^{S (a)},\\
&&\bar{M}_{7\gamma}^{S(b)}(m_s)=-\bar{M}_{7\gamma}^{P(b)}(m_s)
=r_s\frac{\xi_t^*}{\xi_t}M_{7\gamma}^{S (b)}.
\label{3.19}
\end{eqnarray}

In ${\bar{B}}$
meson decay,
the relations ${\bar{M}_{7\gamma}^S(m_b)=\bar{M}_{7\gamma}^P(m_b)}$
and ${\bar{M}_{7\gamma}^S(m_s)=-\bar{M}_{7\gamma}^P(m_s)}$
are satisfied,
and they indicate that, in ${\bar{B}}$
meson decay, the photon chiralities are
left-handed and right-handed, respectively.

We can summarize
the ${B}$ and ${\bar{B}}$
meson decay amplitudes ${A_R}$, ${A_L}$, ${\bar{A}_R}$,and ${\bar{A}_L}$
caused by ${O_{7\gamma}}$ operator as follows:
\begin{eqnarray}
A_{R_{7\gamma}}&=&\left(M_{7\gamma}^{S(a)}(m_b)+M_{7\gamma}^{S(b)}(m_b)\right)
\left[(\epsilon_{\gamma}\cdot \epsilon_{K^*})-i\epsilon_{\mu\nu +-}
\epsilon_{\gamma}^{\mu}\epsilon_{K^*}^{\nu}
\right],
\label{56}\\
A_{L_{7\gamma}}&=&\left(M_{7\gamma}^{S(a)}(m_s)+M_{7\gamma}^{A(b)}(m_s)\right)
\left[(\epsilon_{\gamma}\cdot \epsilon_{K^*})+i\epsilon_{\mu\nu +-}
\epsilon_{\gamma}^{\mu}\epsilon_{K^*}^{\nu}
\right],\label{57}\\
\bar{A}_{R_{7\gamma}}&=&-\frac{\xi_t^*}{\xi_t}\left(M_{7\gamma}^{S(a)}(m_s)
+M_{7\gamma}^{S(b)}(m_s)\right)
\left[(\epsilon_{\gamma}\cdot \epsilon_{K^*})-i\epsilon_{\mu\nu +-}
\epsilon_{\gamma}^{\mu}\epsilon_{K^*}^{\nu}
\right],\label{58}\\
\bar{A}_{L_{7\gamma}}&=&-\frac{\xi_t^*}{\xi_t}\left(M_{7\gamma}^{S(a)}(m_b)
+M_{7\gamma}^{S(b)}(m_b)\right)
\left[(\epsilon_{\gamma}\cdot \epsilon_{K^*})+i\epsilon_{\mu\nu +-}
\epsilon_{\gamma}^{\mu}\epsilon_{K^*}^{\nu}
\right].\label{59}
\end{eqnarray}
From  Eqs.(\ref{56}), (\ref{57}), (\ref{58}), 
and (\ref{59}),
we can see that
${A_{L_{7\gamma}}}$ and ${\bar{A}_{R_{7\gamma}}}$ 
are certainly suppressed compared to
${A_{R_{7\gamma}}}$ and ${\bar{A}_{L_{7\gamma}}}$
 by the factor of ${m_s/m_b}$ within the SM.
Furthermore, we can see the relationships
${A_{R_{7\gamma}}/\xi_t=-\bar{A}_{L_{7\gamma}}/\xi_t^*}$
and ${A_{L_{7\gamma}}/\xi_t=-\bar{A}_{R_{7\gamma}}/\xi_t^*}$
explicitly, which are discussed in Eqs.(\ref{2.28}).

\subsection{${O_{8g}}$ contribution}

\begin{figure}
\begin{center}
\includegraphics[width=4.2cm]{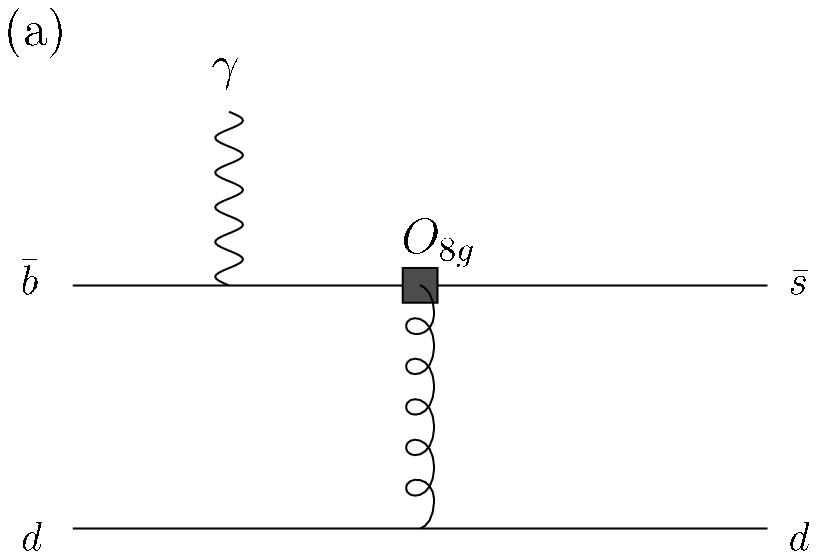}
\hspace{1cm}
\includegraphics[width=4.2cm]{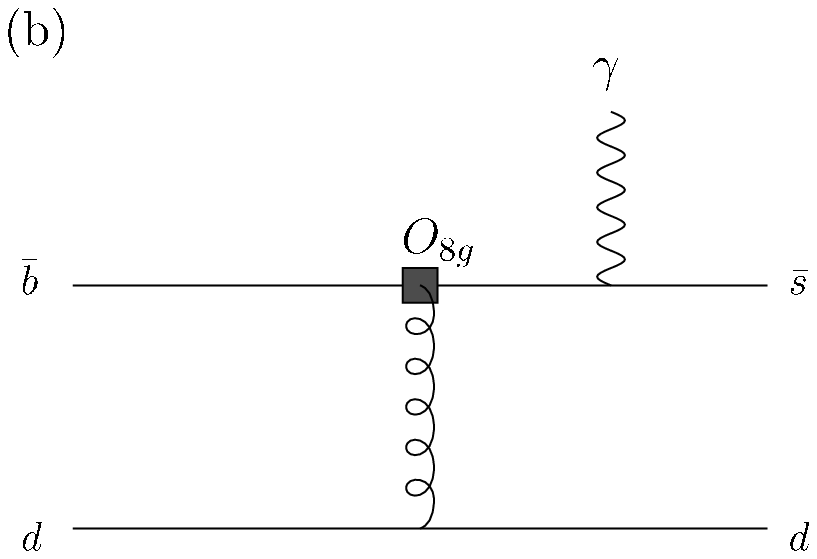}
\vspace{5mm}

\includegraphics[width=4.2cm]{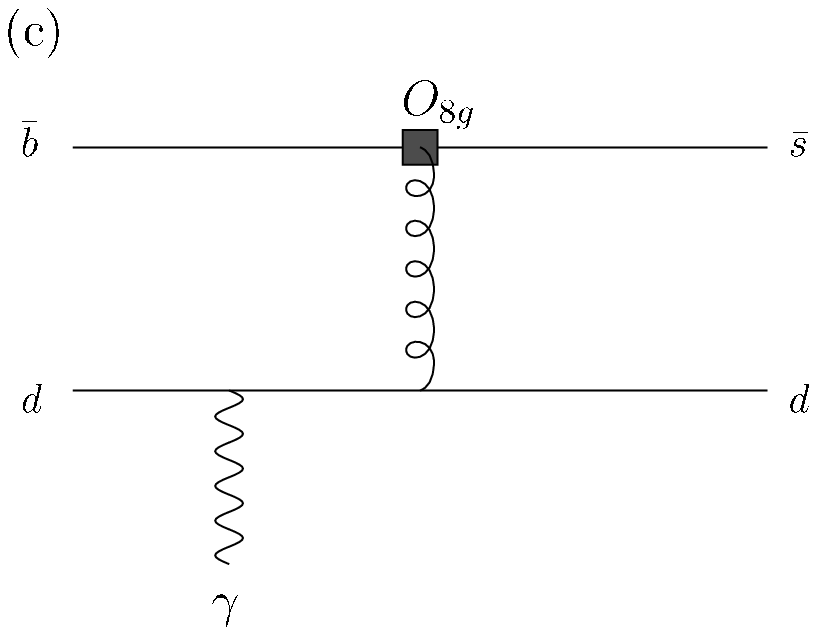}
\hspace{1cm}
\includegraphics[width=4.2cm]{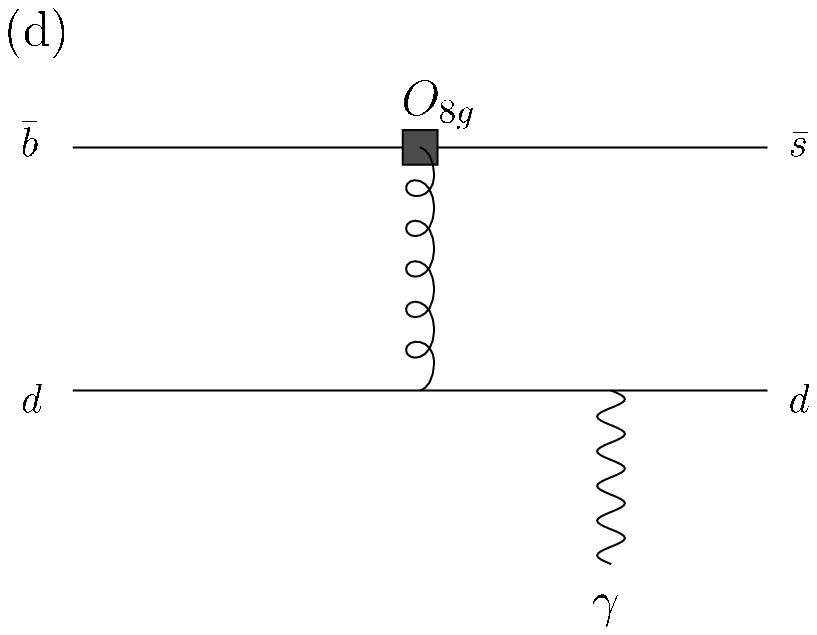}
 \caption{Diagrams for the contribution of operator $O_{8g}$.
 The hard gluon is emitted from the operator
and glued to the spectator quark,
and photon is emitted by the bremsstrahlung
from the external quark line.}
 \label{O8g}
  \end{center}
\end{figure}

Next we concentrate on ${O_{8g}}$ contributions
as shown in Fig.\ref{O8g}.
From Eqs.(19)-(22)
in \cite{Keum:2004is},
we can see that the amplitudes from
${M_{8g}^{(a)}(m_b)}$ to ${ M_{8g}^{(c)}(m_b)}$ satisfy
the relationship ${M_{8g}^S(m_b)=-M_{8g}^P(m_b)}$,
and they imply that the chirality of the photon
from the ${B}$ meson decay 
caused by the component of the ${O_{8g}}$
operator which is proportional to ${m_b}$
is right-handed
except for ${M_{8g}^{(d)}(m_b)}$.
When we concentrate on the exceptional case
${M_{8g}^{(d)}}$, on the other hand,
the twist-2 (leading) component for the ${K^*}$
meson wave function ${\phi_{K^*}^T}$ also has
 ${M_{8g}^S(m_b)=-M_{8g}^P(m_b)}$ relation.
The origins to generate the difference
between ${M_{8g}^S(m_b)}$ and ${M_{8g}^P(m_b)}$
are higher twist components, 
then the effect of the anomaly of the photon chirality
should be  small.
If we divide ${M_{8g}^{(d)}(m_b)}$ components
into right-handed and left-handed photon chirality amplitudes
${M_{8g}^{(d)R}(m_b)}$ and ${M_{8g}^{(d)L}(m_b)}$
by using the relationship as
\begin{eqnarray}
M_{8g}^{(d)}(m_b)&=&(\epsilon_{\gamma}\cdot
 \epsilon_{K^*})M_{8g}^{S(d)}(m_b)
+i\epsilon_{\mu\nu +-}\epsilon_{\gamma}^{\mu}
\epsilon_{K^*}^{\nu}M_{8g}^{P(d)}(m_b)\nonumber\\
&=&M_{8g}^{S(d)R}(m_b)\left[(\epsilon_{\gamma}\cdot
 \epsilon_{K^*})-i\epsilon_{\mu\nu +-}\epsilon_{\gamma}^{\mu}
\epsilon_{K^*}^{\nu}\right]
+M_{8g}^{S(d)L}(m_b)\left[(\epsilon_{\gamma}\cdot
 \epsilon_{K^*})+i\epsilon_{\mu\nu +-}\epsilon_{\gamma}^{\mu}
\epsilon_{K^*}^{\nu}\right],\nonumber\\
\label{60}
\end{eqnarray}
they can be expressed as follows:
\begin{eqnarray}
M_{8g}^{S(d)R}(m_b)&=&-M_{8g}^{P(d)R}(m_b)\nonumber\\
&=&-F^{(0)}Q_d\xi_t r_b
\int^1_0 dx_1 dx_2 \int b_1 d_1 \hspace{1mm}b_2 db_2
\hspace{1mm}\phi_B(x_1,b_1)\hspace{1mm}S_t(x_2)\hspace{1mm}
\alpha_s(t_8^d)\hspace{1mm}
e^{\left[-S_B(t_8^d)-S_{K^{\ast}}(t_8^d)\right]}\nonumber\\
&\times &C_8(t_8^d)
\left[3x_2
r_{K^{\ast}}[\phi_{K^{\ast}}^v(x_2)+\phi_{K^{\ast}}^a(x_2)]
+(2+x_2-x_1)\phi_{K^{\ast}}^T(x_2)\right]
\nonumber\\
&\times & 
H_8^{(d)}
(\sqrt{|A_8'^2|}b_1,E_8b_1,E_8b_2),
\end{eqnarray}
\begin{eqnarray}
M_{8g}^{S(d)L}(m_b)&=&M_{8g}^{P(d)L}(m_b)\nonumber\\
&=&-F^{(0)}Q_d\xi_t r_b
\int^1_0 dx_1 dx_2 \int b_1 d_1 \hspace{1mm}b_2 db_2
\hspace{1mm}\phi_B(x_1,b_1)\hspace{1mm}S_t(x_2)\hspace{1mm}
\alpha_s(t_8^d)\hspace{1mm}
e^{\left[-S_B(t_8^d)-S_{K^{\ast}}(t_8^d)\right]}\nonumber\\
&\times &C_8(t_8^d)
3x_2r_{K^{\ast}}\left[\phi_{K^{\ast}}^v(x_2)-\phi_{K^{\ast}}^a(x_2)\right]
H_8^{(d)}
(\sqrt{|A_8'^2|}b_1,E_8b_1,E_8b_2).
\end{eqnarray}
Then the decay amplitudes with left and right photon chirality
can be expressed as follows:

\begin{eqnarray}
A_{R_{8g}}&=&\left(M_{8g}^{S(a)}(m_b)+M_{8g}^{S(b)}(m_b)
+M_{8g}^{S(c)}(m_b)+M_{8g}^{S(d)R}(m_b)+M_{8g}^{S(d)R}(m_s)\right)\nonumber\\
&&\times 
\left[(\epsilon_{\gamma}\cdot \epsilon_{K^*})-i\epsilon_{\mu\nu +-}
\epsilon_{\gamma}^{\mu}\epsilon_{K^*}^{\nu}
\right],
\label{84}\\
A_{L_{8g}}&=&
\left(M_{8g}^{S(a)}(m_s)+M_{8g}^{S(b)}(m_s)
+M_{8g}^{S(c)}(m_s)+M_{8g}^{S(d)L}(m_s)+M_{8g}^{S(d)L}(m_b)\right)\nonumber\\
&&\times 
\left[(\epsilon_{\gamma}\cdot \epsilon_{K^*})+i\epsilon_{\mu\nu +-}
\epsilon_{\gamma}^{\mu}\epsilon_{K^*}^{\nu}
\right],\\
\label{85}
\bar{A}_{R_{8g}}&=&-\frac{\xi_t^*}{\xi_t}
\left(M_{8g}^{S(a)}(m_s)+M_{8g}^{S(b)}(m_s)
+M_{8g}^{S(c)}(m_s)+M_{8g}^{S(d)L}(m_s)+M_{8g}^{S(d)L}(m_b)\right)\nonumber\\
&&\times 
\left[(\epsilon_{\gamma}\cdot \epsilon_{K^*})-i\epsilon_{\mu\nu +-}
\epsilon_{\gamma}^{\mu}\epsilon_{K^*}^{\nu}
\right],\\
\label{86}
\bar{A}_{L_{8g}}&=&-\frac{\xi_t^*}{\xi_t}
\left(M_{8g}^{S(a)}(m_b)+M_{8g}^{S(b)}(m_b)
+M_{8g}^{S(c)}(m_b)+M_{8g}^{S(d)R}(m_b)+M_{8g}^{S(d)R}(m_s)\right)\nonumber\\
&&\times 
\left[(\epsilon_{\gamma}\cdot \epsilon_{K^*})+i\epsilon_{\mu\nu +-}
\epsilon_{\gamma}^{\mu}\epsilon_{K^*}^{\nu}
\right].
\label{87}
\end{eqnarray}
In this way, we can classify all the decay amplitudes
in \cite{Keum:2004is}
into ones
with left-handed or right-handed photon chiralities.
\subsection{Loop contributions}
\subsubsection{Quark line photon emission}

\begin{figure}
\begin{center}
\includegraphics[width=4.2cm]{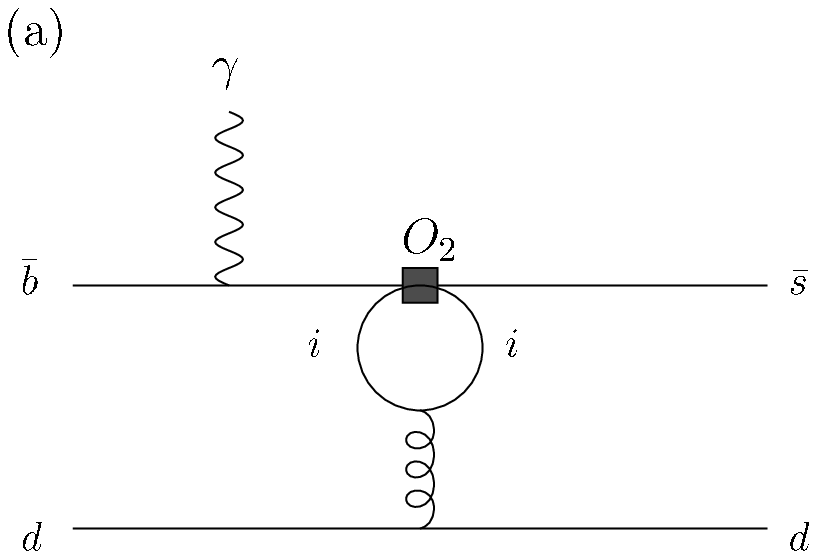}
\hspace{1cm}
\includegraphics[width=4.2cm]{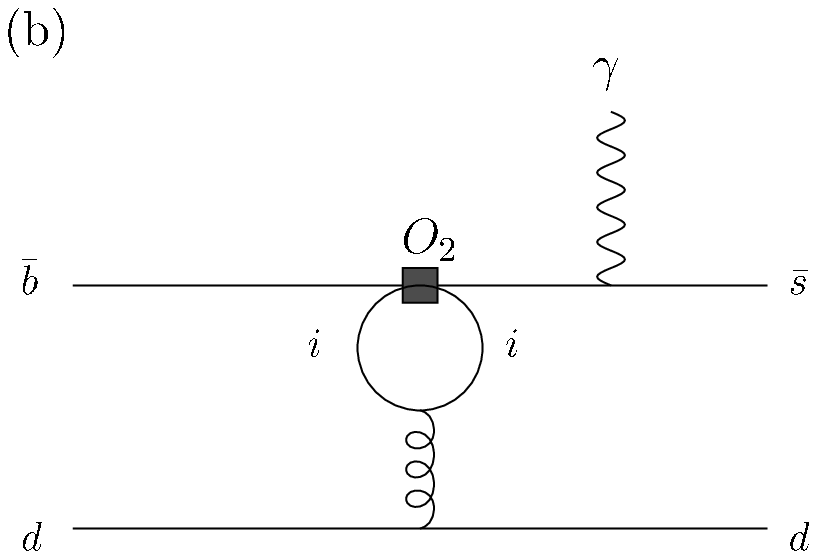}
\vspace{5mm}

\includegraphics[width=4.2cm]{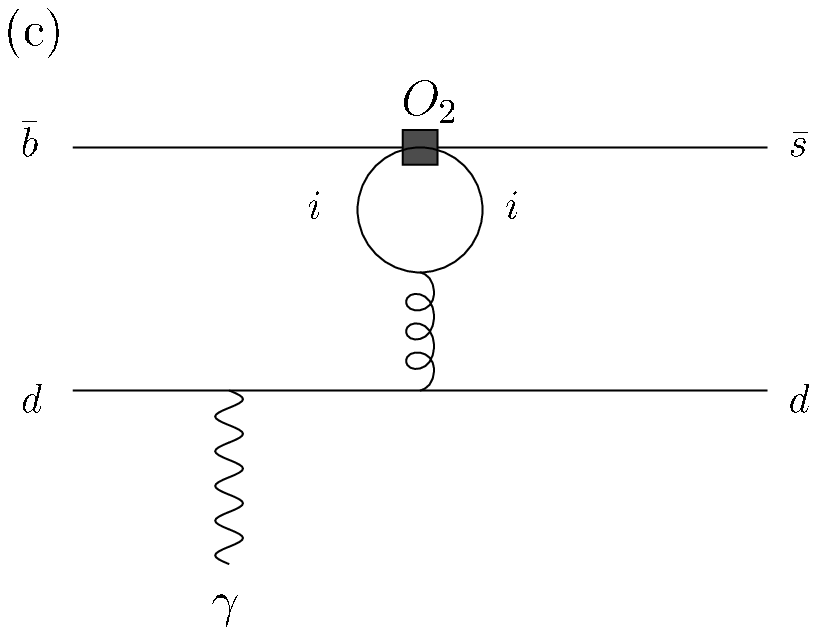}
\hspace{1cm}
\includegraphics[width=4.2cm]{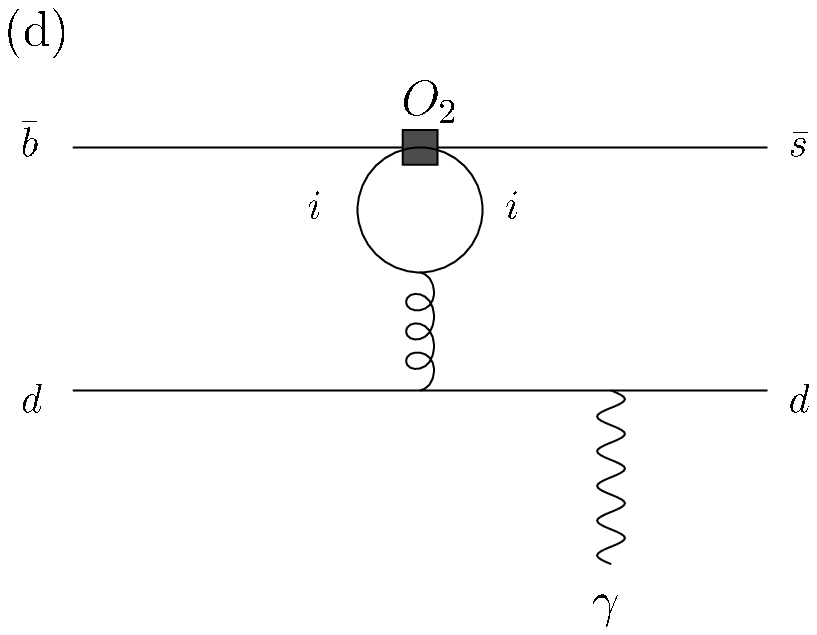}
 \caption{Diagrams for ${i=u,c}$ 
quark loop contributions inserted
 $O_2$ operator, with the
 photon emitted from the external quark line.}
 \label{loopa}
\end{center}
\end{figure}

Here we mention about the charm and up loop quark contribution 
as in Fig.\ref{loopa}.
In this case, the photon is emitted through the external quark lines.
The detailed expressions for the decay amplitudes are 
shown from Eqs.(31)-(35)
in
\cite{Keum:2004is},
and ${M_{1i}^{S(a)}=M_{1i}^{P(a)}}$ means the photon chirality
is left-handed, and
${M_{1i}^{S(b)}=-M_{1i}^{P(b)}}$ and ${M_{1i}^{S(c)}=-M_{1i}^{P(c)}}$
indicate that the photon chirality
is right-handed.
In the ${M_{1i}^{(d)}}$ case,
we can divide the amplitudes into the ones
with left-handed or right-handed
photon chiralities by using Eq.(\ref{60}):
\begin{eqnarray}
M_{1i}^{S(d)R}&=&-M_{1i}^{S(d)R}\nonumber\\
&=&\frac{Q_d}{2}F^{(0)}\hspace{1mm}\xi_i
\int^1_0 dx_1 dx_2 \int b_1 db_1 b_2 db_2 
\hspace{1mm}
e^{\left[-S_B(t_2^d)-S_{K^*}(t_2^d)\right]}C_2(t_2^d)\phi_B(x_1,b_1)
\nonumber\\
&\times &\alpha_s(t_2^d)S_t(x_2)
H_2^{(d)}(\sqrt{|A_2^{'2}|}b_1,E_2b_1,E_2b_2)
\Big[G(m_i^2,-A_2^{'2},t_2^d)
-\frac{2}{3}\Big]
\nonumber\\
&
\times &
\left[
x_2r_{K^*}(1+2x_2)[\phi_{K^*}^v(x_2)+\phi_{K^*}^a(x_2)]
+3(x_2-x_1)\phi_{K^*}^T(x_2)
\right],
\end{eqnarray}
\begin{eqnarray}
M_{1i}^{S(d)L}&=&M_{1i}^{S(d)L}\nonumber\\
&=&\frac{Q_d}{2}F^{(0)}\hspace{1mm}\xi_i
\int^1_0 dx_1 dx_2 \int b_1 db_1 b_2 db_2 
\hspace{1mm}
e^{\left[-S_B(t_2^d)-S_{K^*}(t_2^d)\right]}C_2(t_2^d)\phi_B(x_1,b_1)
\nonumber\\
&\times &\alpha_s(t_2^d)S_t(x_2)
H_2^{(d)}(\sqrt{|A_2^{'2}|}b_1,E_2b_1,E_2b_2)
\Big[G(m_i^2,-A_2^{'2},t_2^d)
-\frac{2}{3}\Big]
\nonumber\\
&
\times &
x_2r_{K^*}(2+x_2)\left[\phi_{K^*}^v(x_2)-\phi_{K^*}^a(x_2)\right].
\end{eqnarray}
Then the decay amplitudes with each photon chirality
can be expressed
as follows:
\begin{eqnarray}
A_{R_{1i}}&=&\left(M_{1i}^{S(b)}+M_{1i}^{S(c)}
+M_{1i}^{S(d)R}\right)
\left[(\epsilon_{\gamma}\cdot \epsilon_{K^*})-i\epsilon_{\mu\nu +-}
\epsilon_{\gamma}^{\mu}\epsilon_{K^*}^{\nu}
\right],\\
A_{L_{1i}}&=&\left(M_{1i}^{S(a)}
+M_{1i}^{S(d)L}\right)
\left[(\epsilon_{\gamma}\cdot \epsilon_{K^*})+i\epsilon_{\mu\nu +-}
\epsilon_{\gamma}^{\mu}\epsilon_{K^*}^{\nu}
\right],\\
\bar{A}_{R_{1i}}&=&-\frac{\xi_i^*}{\xi_i}\left(M_{1i}^{S(a)}
+M_{1i}^{S(d)L}\right)\left[(\epsilon_{\gamma}\cdot \epsilon_{K^*})
-i\epsilon_{\mu\nu +-}
\epsilon_{\gamma}^{\mu}\epsilon_{K^*}^{\nu}
\right],\\
\bar{A}_{L_{1i}}&=&-\frac{\xi_i^*}{\xi_i}
\left(M_{1i}^{S(b)}+M_{1i}^{S(c)}
+M_{1i}^{S(d)R}\right)
\left[(\epsilon_{\gamma}\cdot \epsilon_{K^*})+i\epsilon_{\mu\nu +-}
\epsilon_{\gamma}^{\mu}\epsilon_{K^*}^{\nu}
\right].
\end{eqnarray}

\subsubsection{Loop line photon emission}
\begin{figure}
\begin{center}
\includegraphics[width=4.8cm]{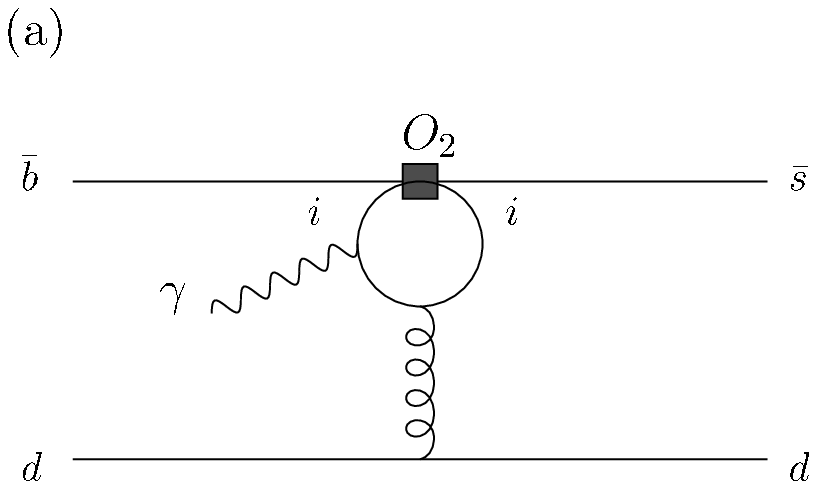}
\hspace{1cm}
\includegraphics[width=4.8cm]{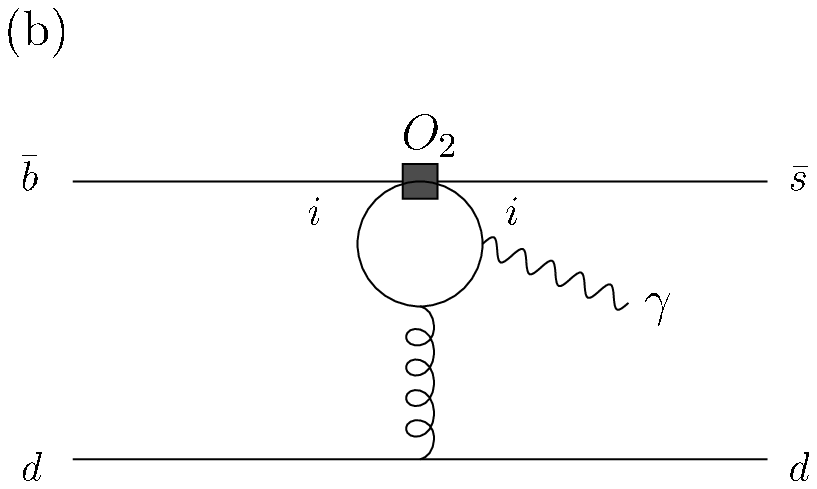}
 \caption{Diagrams for ${i=u,c}$ 
quark loop contributions inserted
 $O_2$ operator, with the
 photon emitted from the internal loop quark line.}
  \label{loopb}
  \end{center}
\end{figure}
Next we consider the loop contributions
in which the photon is emitted from the internal quark loop line
as shown in Fig.\ref{loopb}.
The explicit formulas for the decay amplitudes are 
Eqs.(42) and (43)
in
\cite{Keum:2004is}.
In this case, we can also divide the decay amplitudes
into the ones with the each photon chirality
by using  Eq.(\ref{60}):% as follows: 
\begin{eqnarray}
M_{2i}^{SR}&=&-M_{2i}^{PR}\nonumber\\
&=&-\frac{4}{3}F^{(0)}\hspace{1mm}\xi_i
\int^1_0 dx \int^{1-x}_0 dy
\int^1_0 dx_1 dx_2 \int b_1 db_1  \phi_B(x_1,b_1)
C_2(t_2)\alpha_s(t_2)e^{\left[-S_B(t_2)-S_{K^*}(t_2)
\right]}
\nonumber\\
&
\times & 
\frac{1}{xyx_2 M_B^2-m_i^2}
\Big[xyx_2\left[(1-x_2)r_{K^*}(\phi_{K^*}^v(x_2)+\phi_{K^*}^a(x_2))-(1+2x_1)
\phi_{K^*}^T(x_2)
\right]\nonumber\\
&&+x(1-x)\left[
x_2^2r_{K^*}(\phi_{K^*}^v(x_2)+\phi_{K^*}^a(x_2))+3x_1x_2\phi_{K^*}^T(x_2)
\right]\Big]H_2(b_1A,b_1\sqrt{|B^2|}),
\end{eqnarray}
\begin{eqnarray}
M_{2i}^{SL}&=&M_{2i}^{PL}\nonumber\\
&=&\frac{4}{3}F^{(0)}\hspace{1mm}\xi_i
\int^1_0 dx \int^{1-x}_0 dy
\int^1_0 dx_1 dx_2 \int b_1 db_1  \phi_B(x_1,b_1)
C_2(t_2)\alpha_s(t_2)e^{\left[-S_B(t_2)-S_{K^*}(t_2)
\right]}
\nonumber\\
&
\times &
\frac{1}{xyx_2 M_B^2-m_i^2}~
xyx_2^2r_{K^*}\left[
\phi_{K^*}^v(x_2)-\phi_{K^*}^a(x_2)
\right]H_2(b_1A,b_1\sqrt{|B^2|}),
\label{76}
\end{eqnarray}
and
we can summarize the decay amplitudes 
with each photon chirality as follows:
\begin{eqnarray}
A_{R_{2i}}&=&M_{2i}^{SR}
\left[(\epsilon_{\gamma}\cdot \epsilon_{K^*})-i\epsilon_{\mu\nu +-}
\epsilon_{\gamma}^{\mu}\epsilon_{K^*}^{\nu}
\right],\\
A_{L_{2i}}&=&M_{2i}^{SL}
\left[(\epsilon_{\gamma}\cdot \epsilon_{K^*})+i\epsilon_{\mu\nu +-}
\epsilon_{\gamma}^{\mu}\epsilon_{K^*}^{\nu}
\right],\\
\bar{A}_{R_{2i}}&=&-\frac{\xi_i^*}{\xi_i}M_{2i}^{SL}
\left[(\epsilon_{\gamma}\cdot \epsilon_{K^*})-i\epsilon_{\mu\nu +-}
\epsilon_{\gamma}^{\mu}\epsilon_{K^*}^{\nu}
\right],\\
\bar{A}_{L_{2i}}&=&-\frac{\xi_i^*}{\xi_i}M_{2i}^{SR}
\left[(\epsilon_{\gamma}\cdot \epsilon_{K^*})+i\epsilon_{\mu\nu +-}
\epsilon_{\gamma}^{\mu}\epsilon_{K^*}^{\nu}
\right]. 
\end{eqnarray}

\subsection{Annihilation contributions}

\begin{figure}
\begin{center}
\includegraphics[width=4.2cm]{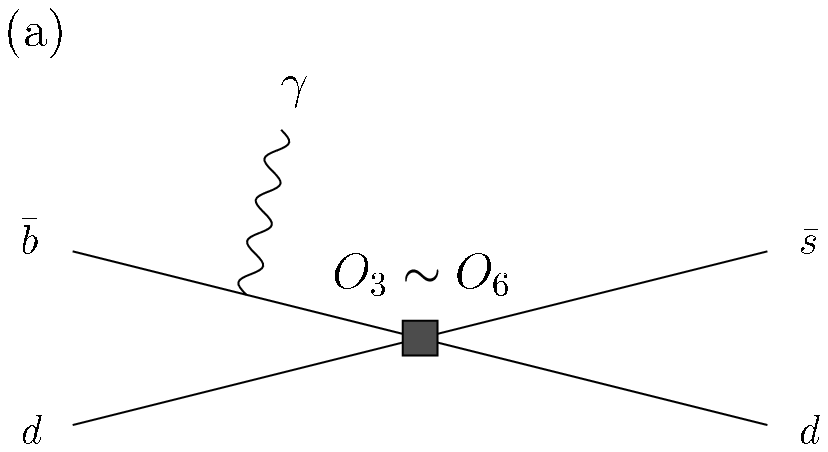}
\hspace{1cm}
\includegraphics[width=4.2cm]{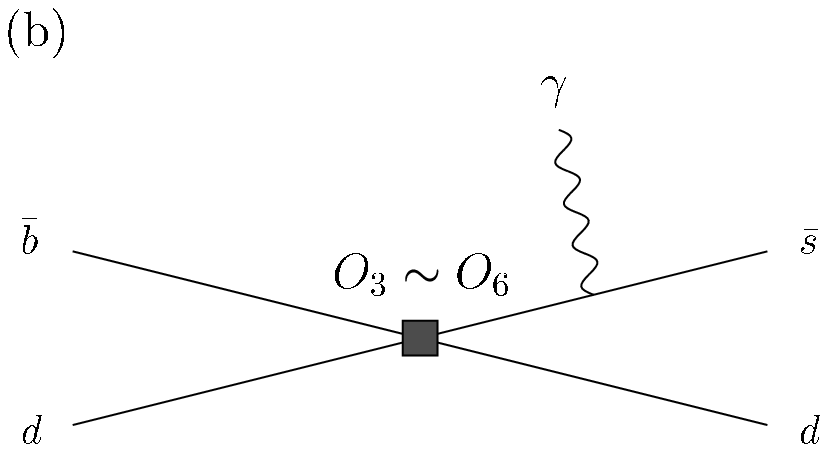}
\vspace{5mm}

\includegraphics[width=4.2cm]{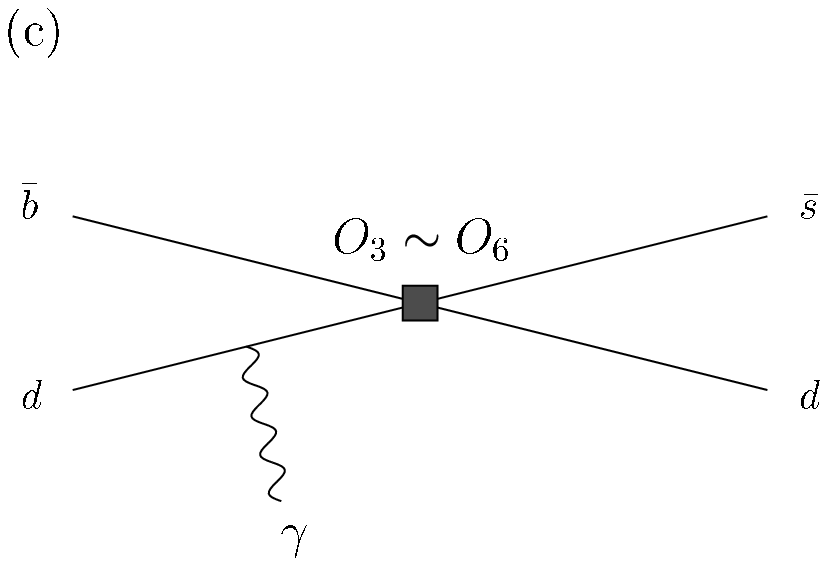}
\hspace{1cm}
\includegraphics[width=4.2cm]{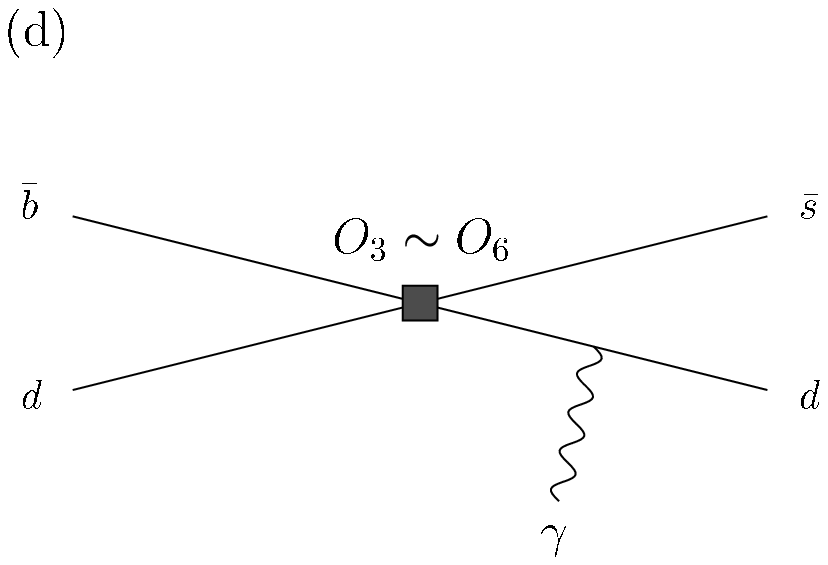}
\caption{Annihilation diagrams with QCD penguin operators $O_i$
 inserted.}
 \label{anni1}
\end{center}
\end{figure}

In this section, we want to discuss  the annihilation
contributions caused by QCD penguin operators from ${O_3}$ to ${O_6}$
shown in
Fig.\ref{anni1}.
In the neutral mode, the tree annihilations caused by
${O_1,~O_2}$ do not exist.
${O_3}$, ${O_4}$, and 
${O_5}$, ${O_6}$ operators
have ${(V-A) (V-A)}$, and ${(V-A)(V+A)}$ vertices, respectively,
and we define ${a_4(t)=C_4(t)+C_3(t)/3,
a_6(t)=C_6(t)+C_5(t)/3}$.
From Eqs.(55)-(62) in
\cite{Keum:2004is},
we can see that ${M_4^{(a)}}$ has left-handed photon chirality,
and ${M_4^{(c)}}$, ${M_6}^{(b)}$, and ${M_{6}^{(d)}}$ have
right-handed photon chirality.
Furthermore, we can divide ${M_4^{(b)}}$ and ${M_4^{(d)}}$
into the ones with left- or right-handed photon chiralities as follows:
\begin{eqnarray}
M_4^{S(b)R}&=&-M_4^{P(b)R}\nonumber\\
&=&F^{(0)}\frac{3\sqrt{6}Q_sf_B\pi}{4M_B^2}r_{K^*}\xi_t
\int^1_0 dx_2 \int b_2 db_2
\hspace{1mm} a_4(t_a^b)S_t(x_2)\hspace{1mm}
e^{\left[-S_{K^*}(t_a^b)\right]}i\frac{\pi}{2}H_0^{(1)}(b_2B_a)\nonumber\\
&
\times &\Big[\phi_{K^*}^{v}(x_2)+ \phi_{K^*}^a(x_2)\Big],
\end{eqnarray}
\begin{eqnarray}
M_4^{S(b)L}&=&M_4^{P(b)L}\nonumber\\
&=&F^{(0)}\frac{3\sqrt{6}Q_sf_B\pi}{4M_B^2}r_{K^*}\xi_t
\int^1_0 dx_2 \int b_2 db_2
\hspace{1mm} a_4(t_a^b)S_t(x_2)\hspace{1mm}
e^{\left[-S_{K^*}(t_a^b)\right]}i\frac{\pi}{2}H_0^{(1)}(b_2B_a)\nonumber\\
&\times &(1-x_2)\Big[\phi_{K^*}^{v}(x_2)+ \phi_{K^*}^a(x_2)\Big].
\end{eqnarray}
\begin{eqnarray}
M_4^{S(d)R}&=&-M_4^{P(d)R}\nonumber\\
&=&-F^{(0)}\frac{3\sqrt{6}Q_sf_B\pi}{4M_B^2}r_{K^*}\xi_t
\int^1_0 dx_2 \int b_2 db_2
\hspace{1mm} a_4(t_a^b)S_t(x_2)\hspace{1mm}
e^{\left[-S_{K^*}(t_a^b)\right]}i\frac{\pi}{2}H_0^{(1)}(b_2B_a)\nonumber\\
&\times &x_2\Big[\phi_{K^*}^{v}(x_2)+ \phi_{K^*}^a(x_2)\Big],
\end{eqnarray}
\begin{eqnarray}
M_4^{S(d)L}&=&M_4^{P(d)L}\nonumber\\
&=&-F^{(0)}\frac{3\sqrt{6}Q_sf_B\pi}{4M_B^2}r_{K^*}\xi_t
\int^1_0 dx_2 \int b_2 db_2
\hspace{1mm} a_4(t_a^b)S_t(x_2)\hspace{1mm}
e^{\left[-S_{K^*}(t_a^b)\right]}i\frac{\pi}{2}H_0^{(1)}(b_2B_a)\nonumber\\
&\times &\Big[\phi_{K^*}^{v}(x_2)- \phi_{K^*}^a(x_2)\Big].
\end{eqnarray}

Then we can summarize the decay amplitudes with
each photon chirality as follows:
\begin{eqnarray}
A_{R_{4}}&=&\left(M_{4}^{S(b)R}
+M_{4}^{S(c)}+M_{4}^{S(d)R}\right)
\left[(\epsilon_{\gamma}\cdot \epsilon_{K^*})-i\epsilon_{\mu\nu +-}
\epsilon_{\gamma}^{\mu}\epsilon_{K^*}^{\nu}
\right],\\
A_{L_{4}}&=&\left(M_{4}^{S(a)}+M_{4}^{S(b)L}
+M_{4}^{S(d)L}\right)
\left[(\epsilon_{\gamma}\cdot \epsilon_{K^*})+i\epsilon_{\mu\nu +-}
\epsilon_{\gamma}^{\mu}\epsilon_{K^*}^{\nu}
\right],\\
\bar{A}_{R_{4}}&=&
-\frac{\xi_t^*}{\xi_t}\left(M_{4}^{S(a)}+M_{4}^{S(b)L}
+M_{4}^{S(d)L}\right)
\left[(\epsilon_{\gamma}\cdot \epsilon_{K^*})-i\epsilon_{\mu\nu +-}
\epsilon_{\gamma}^{\mu}\epsilon_{K^*}^{\nu}
\right],\\
\bar{A}_{L_{4}}&=&-\frac{\xi_t^*}{\xi_t}\left(M_{4}^{S(b)R}
+M_{4}^{S(c)}+M_{4}^{S(d)R}\right)
\left[(\epsilon_{\gamma}\cdot \epsilon_{K^*})+i\epsilon_{\mu\nu +-}
\epsilon_{\gamma}^{\mu}\epsilon_{K^*}^{\nu}
\right],
\end{eqnarray}
\begin{eqnarray}
A_{R_{6}}&=&\left(M_{6}^{S(b)}
+M_{6}^{S(d)}\right)
\left[(\epsilon_{\gamma}\cdot \epsilon_{K^*})-i\epsilon_{\mu\nu +-}
\epsilon_{\gamma}^{\mu}\epsilon_{K^*}^{\nu}
\right],\\
A_{L_{6}}&=&%\\
\bar{A}_{R_{6}}=0,\\
\bar{A}_{L_{6}}&=&-\frac{\xi_t^*}{\xi_t}
\left(M_{6}^{S(b)}
+M_{6}^{S(d)}\right)
\left[(\epsilon_{\gamma}\cdot \epsilon_{K^*})+i\epsilon_{\mu\nu +-}
\epsilon_{\gamma}^{\mu}\epsilon_{K^*}^{\nu}
\right].
\end{eqnarray}

\subsection{Long distance contributions to the photon quark coupling}
\label{Long distance}
\begin{figure}%[htbp]
\begin{center}
\includegraphics[width=5.3cm]{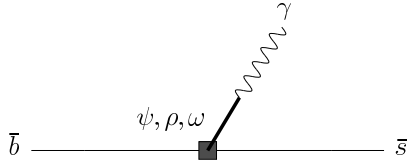}
\caption{Vector-Meson-Dominance contributions
mediated by ${\psi,\rho,\omega}$.}
\label{Vector}
\end{center}
\end{figure}

\begin{figure}
\begin{center}
\includegraphics[width=4.2cm]{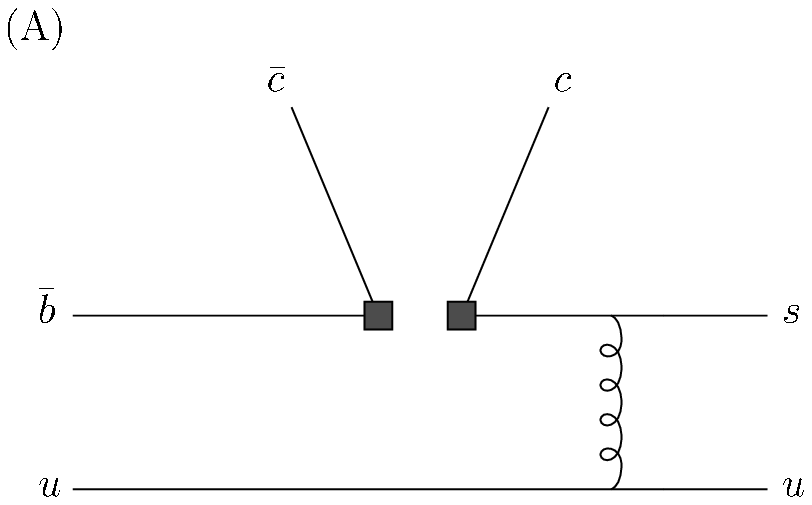}
\hspace{1cm}
\includegraphics[width=4.2cm]{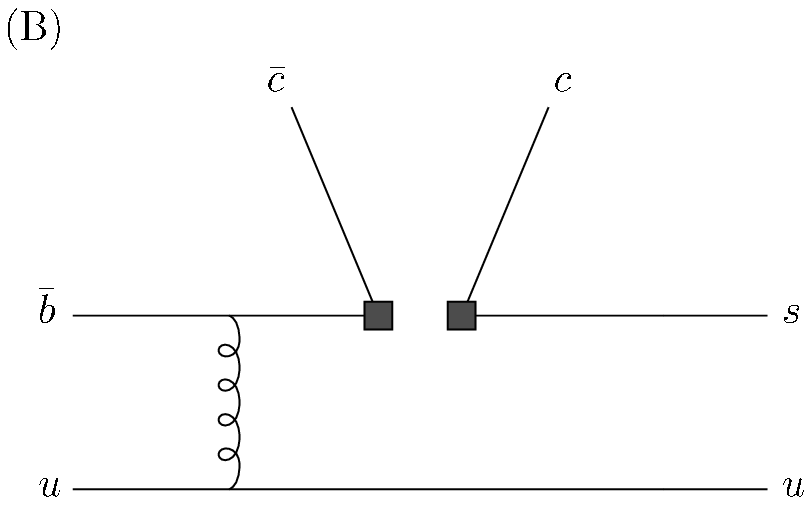}
\vspace{5mm}

\includegraphics[width=4.2cm]{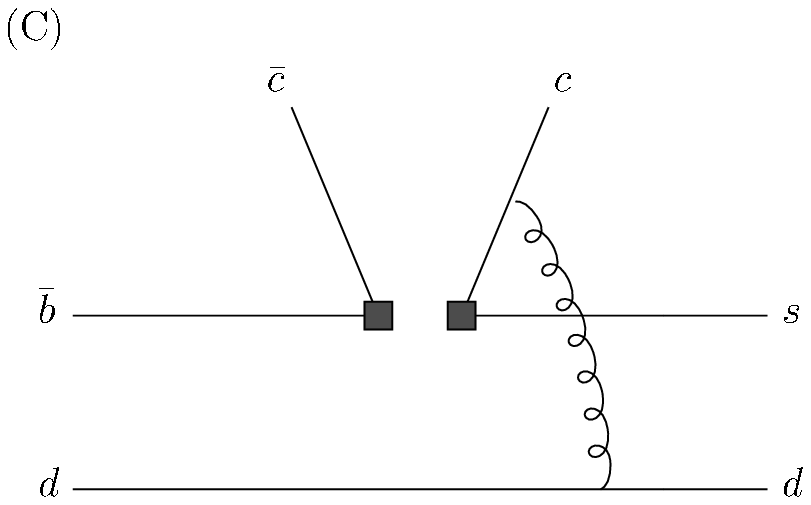}
\hspace{1cm}
\includegraphics[width=4.2cm]{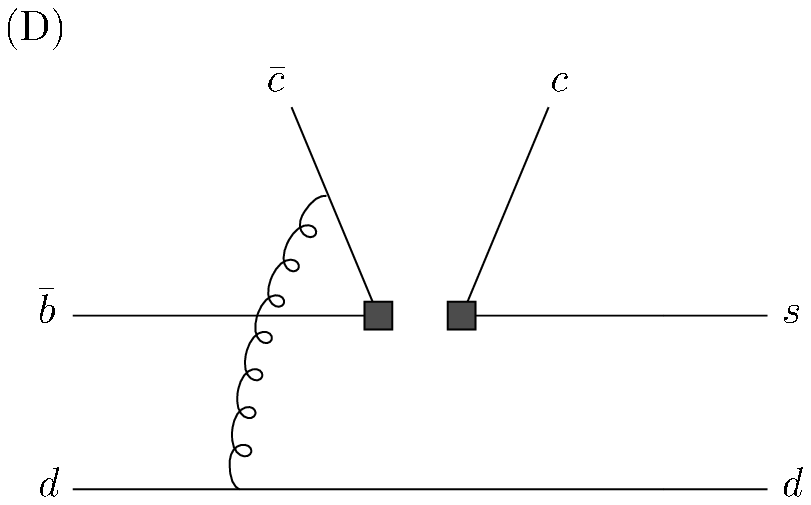}
\caption{(A), (B) are factorizable and (C), (D) are nonfactorizable
 contributions to the hadronic matrix element for 
${<K^*\psi |H_{\mbox{\scriptsize{eff}}}|B>}$. }
\label{non-fact}
\end{center}
\end{figure}

Here we want to discuss the long distance contributions.
In order to examine the SM or search for new physics
indirectly by comparing the experimental data with the
values predicted within the SM, 
we have to take into account these long distance effects:
${B\to K^* (\psi,\rho,\omega) \rightarrow K^* \gamma}$
(Fig.\ref{Vector}).
If we use the vector-meson-dominance,
the ${B\to K^*\gamma}$ decay amplitude can be expressed
as inserting the complete set of possible intermediate
vector meson states like
\begin{eqnarray}
 \langle K^*\gamma | H_{\mbox{\scriptsize{eff}}}|B\rangle
=\sum_{V}\langle\gamma | A_{\nu}J_{\mbox{\scriptsize{em}}}^{\nu}|V\rangle
\frac{-i}{q_V^2-m_V^2}
\langle VK^*  | H_{\mbox{\scriptsize{eff}}}|B\rangle,
\end{eqnarray}
where ${V=\psi,\rho,\omega}$.
Now we concretely consider  ${B\rightarrow K^* \psi \to K^*\gamma}$.
Four diagrams contribute to the
hadronic matrix element of
${\langle K^* \psi  | H_{\mbox{\scriptsize{eff}}}|B\rangle}$ 
(see Fig.\ref{non-fact}),
and first of all, 
we consider the leading contributions:
the factorizable ones shown in Figs. \ref{non-fact}(A) and \ref{non-fact}(B).

\subsubsection{Factorizable contribution}
The detailed explanation for the derivation of the
factorizable 
amplitudes and 
explicit formulas is from Eqs.(77)-(79)
in \cite{Keum:2004is}.
The momentum of the photon ${q^2=0}$ is smaller than
the threshold mass ${4m_u^2}$, then the 
imaginary part from the decay width
in the propagator of the ${\rho}$ meson should not be generated.
The long distance contribution mediated by the ${\rho}$
meson is much smaller than the ${u}$ quark loop contributions,
the change does not effect  the final numerical results.
We can see that ${M_{LD}^{(A)}}$ has right-handed
photon chirality, and for  ${M_{LD}^{(B)}}$,
it is also possible to
divide it into the ones with left- or right-handed
photon chiralities as mentioned above,
and the results become as follows:
\begin{eqnarray}
M^{S(B)R}_{LD}&=&-M^{P(B)R}_{LD}\nonumber\\
&=&\frac{8\pi^2}{M_B^2}F^{(0)}\int^1_0 dx_1 dx_2
\int b_1db_1 b_2db_2 )\phi_B(x_1,b_1)
S_t(x_2)
\alpha_s(t_7^b)
e^{[-S_B(t_7^b)-S_{K^*}(t_7^b)]}
a_1(t_7^b)\nonumber\\
&\times &H_{7}^{(b)}(A_7b_1,C_7b_1, C_7b_2)
\Big[r_{K^*}\phi_{K^*}^v(x_2)
+\phi_{K^*}^T(x_2)
+r_{K^*}\phi_{K^*}^a(x_2)\Big]\nonumber\\
&\times &\left(
\xi_{c}\frac{2\kappa g_{\psi}(m_\psi^2)^2}{3}+
\xi_{u}\left[
\frac{g_{\omega}(m_{\omega}^2)^2}{6}+
\frac{g_{\rho}(m_{\rho}^2)^2}{2
 %(1-i\Gamma/m_{\rho})
}
\right]\right),
\end{eqnarray}
\begin{eqnarray}
M^{S(B)L}_{LD}&=&M^{P(B)L}_{LD}\nonumber\\
&=&\frac{8\pi^2}{M_B^2}F^{(0)}\int^1_0 dx_1 dx_2
\int b_1db_1 b_2db_2 )\phi_B(x_1,b_1)
S_t(x_2)
\alpha_s(t_7^b)
e^{[-S_B(t_7^b)-S_{K^*}(t_7^b)]}
a_1(t_7^b)\nonumber\\
&\times &H_{7}^{(b)}(A_7b_1,C_7b_1, C_7b_2)
r_{K^*}(1+x_2)\left[
\phi_{K^*}^v(x_2)-\phi_{K^*}^a(x_2)
\right]
\nonumber\\
&\times &\left(
\xi_{c}\frac{2\kappa g_{\psi}(m_\psi^2)^2}{3}+
\xi_{u}\left[
\frac{g_{\omega}(m_{\omega}^2)^2}{6}+
\frac{g_{\rho}(m_{\rho}^2)^2}{2
 %(1-i\Gamma/m_{\rho})
}
\right]\right),
\end{eqnarray}
\begin{eqnarray}
A_{R_{LD}}&=&\left(M_{LD}^{S(A)}
+M_{LD}^{S(B)R}\right)
\left[(\epsilon_{\gamma}\cdot \epsilon_{K^*})-i\epsilon_{\mu\nu +-}
\epsilon_{\gamma}^{\mu}\epsilon_{K^*}^{\nu}
\right],\\
A_{L_{LD}}&=&
M_{LD}^{S(B)L}
\left[(\epsilon_{\gamma}\cdot \epsilon_{K^*})+i\epsilon_{\mu\nu +-}
\epsilon_{\gamma}^{\mu}\epsilon_{K^*}^{\nu}
\right],\\
\bar{A}_{R_{LD}}&=&
-\frac{\xi_i^*}{\xi}M_{LD}^{S(B)L}
\left[(\epsilon_{\gamma}\cdot \epsilon_{K^*})-i\epsilon_{\mu\nu +-}
\epsilon_{\gamma}^{\mu}\epsilon_{K^*}^{\nu}
\right],\\
\bar{A}_{L_{LD}}&=&-\frac{\xi_i^*}{\xi}\left(M_{LD}^{S(A)}
+M_{LD}^{S(B)R}\right)
\left[(\epsilon_{\gamma}\cdot \epsilon_{K^*})+i\epsilon_{\mu\nu +-}
\epsilon_{\gamma}^{\mu}\epsilon_{K^*}^{\nu}
\right].
\end{eqnarray}

\subsubsection{Nonfactorizable contribution}
\label{nonfac}
Next we consider the nonfactorizable long distance contributions
as in Figs.\ref{non-fact}(C) and \ref{non-fact}(D).
In order to estimate the effect 
to the total mixing-induced CP asymmetry,
we roughly estimate the degree of 
nonfactorizable long distance contribution to the factorizable long
distance contribution. 
We have already known that
the theoretical computation which is estimated
by including only the factorizable contribution
and experimental data
for the transversely polarized branching ratio in
${B\to J/\psi K^* }$ decay mode do not agree with
each other \cite{Keum:2004is}.
If we assume that the difference between the experimental value
and theoretical prediction is due to the nonfactorizable amplitude,
we expect that it amounts to about 40\% contribution
to the factorizable amplitude.
We include the nonfactorizable contribution
as the free parameter which satisfies the condition such
that it's contribution amounts to about 40\%
to the factorizable one,
and we numerically compute the effect to the mixing-induced CP
asymmetry as show in Sec.\ref{Numerical}.

\section{Numerical results}
\label{Numerical}
We want to show the numerical analysis in this section.
In the evaluation of the various form factors and amplitudes,
we adopt
${G_F=1.16639\times 10^{-5} \mbox{GeV}^{-2}}$,
leading order strong coupling ${\alpha_s}$ defined at 
the flavor number ${n_f=4}$, the decay constants 
${f_B=190 \mbox{MeV}}$, ${f_{K^*}=226 \mbox{MeV}}$,
and ${f_{K^*}^T=185 \mbox{MeV}}$,
the masses 
${M_B=5.28 \mbox{GeV}}$, ${M_{K^*}=0.892 \mbox{GeV}}$, 
${m_b=4.8 \mbox{GeV}}$,
${m_c=1.2 \mbox{GeV}}$, and
${m_s=0.12 \mbox{GeV}}$ \cite{Gockeler:2005nc}, and
the meson lifetime ${\tau_{B^0}=1.542 \hspace{1mm}\mbox{ps}}$.
Furthermore we used the leading order Wilson coefficients 
\cite{Buchalla:1995vs} and we take the 
${K^*}$, ${\rho}$, and ${\omega}$ meson wave functions up to twist-3.

In order to compute the mixing-induced CP asymmetry,
we clarify the theoretical error for it.
\begin{enumerate}
\item
First if we change the meson wave function parameter
in the range ${\omega_B=(0.40\pm 0.04)}$GeV,
the uncertainty of the mixing-induced CP asymmetry
amounts to about 10\%.
Furthermore, the uncertainties from the quark mass amount
to about 20\%. Then the error from the input parameters is about 20\%.
\item 
We can expect that the uncertainty from the higher order
effects should be largely canceled because
we take the ratio in the computation of the asymmetry.
\item
About the error from the Cabibbo-Kobayashi-Maskawa (CKM) parameters,
we change the ${\bar{\rho}}$, ${\bar{\eta}}$
parameters in the range ${\bar{\rho}=\rho(1-\lambda^2/2)}$
${=0.20\pm 0.09}$
and ${\bar{\eta}=\eta(1-\lambda^2/2)=0.33\pm 0.05}$ \cite{Eidelman:2004wy},
and numerically estimate how the physical quantities
are affected by the change of parameters.
The result is that about 30\% error is generated by 
the uncertainty for the CKM parameters.
\item
Furthermore, we introduce the nonfactorizable long distance 
contribution as mentioned in Sec.\ref{nonfac},
and it can generates about 30\% error for the asymmetry.
\item
Finally, we introduce 100\% hadronic uncertainty for
the ${u}$ quark loop,
and 
the asymmetry changes about 10\%.
\end{enumerate}
In summary, the  total theoretical error amounts to about
50\%  for the mixing-induced CP asymmetry.
 
Then the numerical results for the mixing-induced CP asymmetries
in ${B\to K_S\pi^0\gamma}$ and ${B\to K_L\pi^0\gamma}$ like
Eqs.(\ref{35}) and (\ref{36}) 
which are caused by  ${O_{7\gamma}}$,
${O_{8g}}$, ${c}$ and ${u}$
quark loop contributions, QCD annihilations,
and the 
factorizable and nonfactorizable 
long distance contributions 
become as follows:
\begin{eqnarray}
S_{K_S \pi^0\gamma}=-S_{K_L \pi^0\gamma}
=-(3.5\pm 1.7)\times 10^{-2}.
\label{100}
\end{eqnarray}

\section{Conclusion}
\label{conclusion}
In this paper, we compute the mixing-induced CP asymmetry
in the ${B\to K^*\gamma}$ decay mode with perturbative QCD
approach by including also the small contributions
except for the dominant ${O_{7\gamma}}$ amplitude.
Within the SM, we can roughly estimate
the asymmetry as
${
S_{K_S \pi^0\gamma}^{\mbox{\scriptsize{SM}}}
\simeq -2\frac{m_s}{m_b}\sin{2\phi_1}}$
${=-(2.7\pm 0.9)\times 10^{-2}}$,
when we change the CKM parameters as
${\bar{\rho}=0.20\pm 0.09}$ and ${\bar{\eta}=0.33\pm}$
${0.05}$.
If we compute the asymmetry by taking into  account
for only ${O_{7\gamma}}$, our computation becomes 
${S_{K_S \pi^0\gamma}
=-(2.7\pm 0.9)\times 10^{-2}.}$
Comparing the above two values,
we can check that our computation goes well,
and contributions except for ${O_{7\gamma}}$
generate about 15\% asymmetry.
Then we  investigated
what contribution except for ${O_{7\gamma}}$
makes a difference from  Eq.(\ref{100}).
If we neglect the QCD annihilation contributions,
the asymmetry becomes as
${S_{K_S \pi^0\gamma}
=-(3.8\pm 1.7)\times 10^{-2},}$
by abandoning all the long distance contributions,
${S_{K_S \pi^0\gamma}
=-(3.4\pm 1.3)\times 10^{-2},}$
and we discard the ${u}$ and ${c}$ quark loop contributions,
${S_{K_S \pi^0\gamma}=-(3.0\pm 1.4)\times 10^{-2}}$ and
${S_{K_S \pi^0\gamma}=-(3.3\pm 1.6)}$
${\times 10^{-2}}$, respectively.
From the above consideration,
the most effective contribution
which generates the difference
from the prediction 
with only ${O_{7\gamma}}$ amplitude
is the ${u}$ quark loop contribution,
because the amplitude with left-handed photon chirality
for ${B^0}$ meson decay in Eq.(\ref{76})
can interfere  
with the right-handed photon chirality as mentioned in 
\cite{Grinstein:2004uu,Grinstein:2005nu}. 

The experimental data for the mixing-induced CP asymmetry
in ${B\to K_s\pi^0\gamma}$ are given by
Belle and BaBar as follows:
\begin{equation}
S_{K^*\gamma\to K_S\pi^0\gamma}^{\mbox{\scriptsize{ex}}}=
\begin{cases}
-0.79^{+0.63}_{-0.50}\pm 0.10 & \mbox{\cite{Ushiroda:2005sb,Abe:2004xp}},\\
-0.21\pm 0.40 \pm 0.05 & \mbox{\cite{Aubert:2005bu}}.
\end{cases}
\label{106}
\end{equation}
Comparing our prediction in Eq.(\ref{100})
with the experimental data in Eqs.(\ref{106}),
we can see that our theoretical prediction 
is included in the range
of the 
experimental data.
Furthermore, the time-dependent oscillations in ${B\to K_S\pi^0\gamma}$
do not depend on the resonance structure;
whether the ${K_S\pi^0\gamma}$ final state comes from
${K^*}$ or not \cite{Atwood:2004jj}. The fact will be helpful experimentally
to accumulate higher statistics.
Then
we look forward to the improvement of the
experimental error in the near future
and also for the Super B factory when it is built.

\section*{Acknowledgments}
We acknowledge 
helpful and valuable
suggestions, discussions, and advices
with Professor Amarjit Soni and
Professor
Hsiang-nan Li.
A.I.S acknowledges support from JSPS Grant No.C-17540248.

\end{document}